\documentclass[a4paper]{report}
\usepackage[utf8]{inputenc}
\usepackage[T1]{fontenc}
\usepackage{RJournalnew}
\usepackage{amsmath,amssymb,array}
\usepackage{booktabs}

\providecommand{\tightlist}{%
  \setlength{\itemsep}{0pt}\setlength{\parskip}{0pt}}

\usepackage{longtable}

\newlength{\cslhangindent}
\newlength{\csllabelwidth}
\newlength{\cslentryspacingunit} 
  {}%
  {\par}
\newenvironment{CSLReferences}[2] 
 {
  \setlength{\parindent}{0pt}
  \ifodd #1
  \let\oldpar\par
  \def\par{\hangindent=\cslhangindent\oldpar}
  \fi
  \setlength{\parskip}{#2\cslentryspacingunit}
 }%
 {}
\usepackage{calc}

\usepackage{amsmath, amsthm, amsfonts, amssymb, bm, natbib}

\newcommand{\proglang}[1]{\text{#1}}

\newcommand{\bS}{\mbox{\boldmath $S$}}

\newcommand{\bomega}{\mbox{\boldmath $\bomega$}}

\newcommand{\bc}{\begin{center}}
\newcommand{\ec}{\end{center}}

\newcommand{\bi}{\begin{itemize}}
\newcommand{\ei}{\end{itemize}}

\newcommand{\be}{\begin{enumerate}}
\newcommand{\ee}{\end{enumerate}}

\newcommand{\bs}{\begin{slide*}}
\newcommand{\es}{\end{slide*}}

\usepackage{scalefnt}
\usepackage{bm}

\DefineVerbatimEnvironment{Highlighting}{Verbatim}{commandchars=\\\{\}}
\newenvironment{Shaded}{}{}

\newcommand{\AttributeTok}[1]{#1}

\newcommand{\ConstantTok}[1]{#1}
\newcommand{\ControlFlowTok}[1]{\textcolor[rgb]{0.00,0.00,1.00}{#1}}

\newcommand{\DecValTok}[1]{#1}

\newcommand{\FloatTok}[1]{#1}
\newcommand{\FunctionTok}[1]{#1}

\newcommand{\NormalTok}[1]{#1}

\newcommand{\OtherTok}[1]{\textcolor[rgb]{1.00,0.25,0.00}{#1}}

\newcommand{\SpecialCharTok}[1]{\textcolor[rgb]{0.00,0.50,0.50}{#1}}

\newcommand{\StringTok}[1]{\textcolor[rgb]{0.00,0.50,0.50}{#1}}

\usepackage{longtable,booktabs,array}
\usepackage{calc} 
\usepackage{etoolbox}

\begin{document}

\sectionhead{ }
\volume{ }
\volnumber{ }
\year{ }
\month{ }

\begin{article}
\title{Space-Time Smoothing of Survey Outcomes using the R Package SUMMER}

\author{by Zehang Richard Li, Bryan D Martin, Tracy Qi Dong, Geir-Arne Fuglstad, Jessica Godwin, John Paige, Andrea Riebler, Samuel J Clark, and Jon Wakefield}

\maketitle

\abstract{%
The increasing availability of complex survey data, and the continued need for estimates of demographic and health indicators at a fine spatial and temporal scale, has led to the need for spatio-temporal smoothing methods that acknowledge the manner in which the data were collected. The open source R package SUMMER implements a variety of methods for spatial or spatio-temporal smoothing of survey outcomes. In this paper, we focus primarily on demographic and health indicators. Our methods are particularly useful for data from Demographic Health Surveys (DHS) and Multiple Indicator Cluster Surveys (MICS). We build upon functions within the survey package, and use INLA for fast Bayesian computation. This paper includes a brief overview of these methods and illustrates the workflow of processing surveys, fitting space-time smoothing models for both binary and composite indicators, and visualizing results with both simulated data and DHS surveys.
}

\hypertarget{introduction}{%
\section{Introduction}\label{introduction}}

A wealth of health and demographic indicators are now collected across the world, and interest often focuses on patterns in space and time. Spatial patterns indicate potential disparities while temporal trends are important for determining the impact of interventions and to assess whether targets, such as the sustainable development goals (SDGs) are being met (MacFeely 2020). In low- and middle-income countries (LMIC) the most reliable data with sufficient spatial resolution are often collected under complex sampling designs. Common sources of data include the Demographic Health Surveys (DHS) and Multiple Indicator Cluster Surveys (MICS), both of which use multi-stage cluster sampling. A two-stage cluster design is most common for these surveys. A sampling frame of \emph{clusters} (for example, enumeration areas) is constructed, often from a census, and then strata are formed. The strata consist of some administrative geographical partition crossed with urban/rural (with countries having their own definitions of this dichotomy). Then a pre-specified number of clusters are sampled from these strata under some probabilistic scheme, for example, with probability proportional to size (PPS). Different surveys are powered to different geographical levels. Then, within the selected clusters, households are randomly sampled and individuals are sampled within these households, and asked questions on a range of health and demographic variables. This data collection process that must be acknowledged in the analysis to reduce bias and obtain proper uncertainty measures in the prevalence estimates.

Various packages are available for within R for small area estimation (SAE) of prevalence, including the \CRANpkg{sae} package (Molina and Marhuenda 2015) that supports the popular book of Rao and Molina (2015) and includes the famous Fay and Herriot (1979) model and spatial smoothing options. Other packages include \CRANpkg{rsae} (Schoch 2014), \CRANpkg{hbsae} (Boonstra 2012), \CRANpkg{BayesSAE} (Shi 2018) and \CRANpkg{msae} (Permatasari and Ubaidillah 2021).
A more comprehensive list of related packages are described at \ctv{OfficialStatistics}. Most of the existing packages focus on classical SAE models and provide very limited options for fitting spatial and space-time smoothing models.

In this paper we introduce the R package, \CRANpkg{SUMMER}\footnote{The name originally arises from `Spatio-temporal Under-five Mortality Methods for Estimation in R'. As the package becomes a more general toolkit, it now stands for `Sae Unit/area Models and Methods for Estimation in R'}. This package and its details are available on CRAN. \CRANpkg{SUMMER} provides a computational framework and a collection of tools for smoothing and mapping the prevalence of indicators with complex survey data over space and time, with a special focus on estimating mortality rates. Smoothing is important to avoid unstable estimates and combine information from multiple surveys over time. Originally developed for small area estimation of the under-5 child mortality rate (U5MR), the \CRANpkg{SUMMER} package has been extended to broader mortality rate estimation and more general tasks in SAE. The implemented methods have already been successfully applied to a range of data, e.g., subnational estimates of mortality rates (Mercer et al. 2015; Li et al. 2019; Schlüter and Masquelier 2021; Fuglstad, Li, and Wakefield 2021), HIV prevalence (Wakefield, Okonek, and Pedersen 2020) and vaccination coverage (T. Q. Dong and Wakefield 2021). Recently, the \CRANpkg{SUMMER} package was used to obtain the official United Nations Inter-Agency Group for Mortality Estimation (UN IGME) yearly estimates (1990--2021) of U5MR at administrative level \(2\) below the national level (admin-2 estimates) for \(31\) countries in Africa and Asia (United Nations Inter-agency Group for Child Mortality Estimation 2023). Previously, the UN IGME only produced national estimates using the \emph{B3 model} (Alkema and New 2014). The results of these endeavors are available online at \url{https://childmortality.org}.

The main focus of this paper is to provide an overview of the different prevalence models using survey data and how they can be implemented in \CRANpkg{SUMMER}. The rest of the paper is organized as follows. We first briefly describe different methods to estimate prevalence using survey data. We start with a generic binary indicator and proceed with estimating mortality rates. We then provide an overview of the \CRANpkg{SUMMER} package and the workflows of using \CRANpkg{SUMMER} for prevalence mapping. We then discuss three examples for spatial and space-time smoothing of binary and composite indicators with increasing complexity. The first two examples uses simulated data that are included in the \CRANpkg{SUMMER} package. The last example uses the most recent DHS survey from Malawi. Then we illustrate various visualization and model checking tools in the \CRANpkg{SUMMER} package. Finally we conclude with future work.

\hypertarget{space-time-smoothing-using-complex-survey-data}{%
\section{Space-time smoothing using complex survey data}\label{space-time-smoothing-using-complex-survey-data}}

In this section we review different methods to estimate the prevalence of a health outcome from complex survey data. We begin by discussing design-based, \emph{direct} estimates (Rao and Molina 2015) which are based on response data from that area only. Next, we describe space-time smoothing of the direct estimates using a Fay-Herriot model (Fay and Herriot 1979). We discuss both estimating the prevalence of a single binary indicator and the composite indicators such as U5MR. We then describe a cluster-level model to estimate prevalence at finer spatial and temporal resolutions.

\hypertarget{estimating-the-prevalence-of-a-generic-binary-indicator}{%
\subsection{Estimating the prevalence of a generic binary indicator}\label{estimating-the-prevalence-of-a-generic-binary-indicator}}

Consider a study region that is partitioned into \(n\) areas, with interest focusing on estimating the prevalence of a binary indicator in each area, possibly over time. The data are collected via some complex survey design. For each individual \(j\), let \(y_j\) denote the individual's outcome, and \(w_j\) denote the design weight associated with this individual. Further, let \(s_{it}\) represent the indexes of individuals sampled in area \(i\) and in time period \(t\). The design-based estimator (Horvitz and Thompson 1952; Hájek 1971) is
\begin{equation}
\hat p_{it}^{\texttt{HT}} = \frac{\sum_{j \in s_{it}} w_{j}y_j}{\sum_{j \in s_{it}} w_{j}}.
\end{equation}
This is an example of a direct estimate.
The variance of \(\hat p_{it}^{\texttt{HT}}\) can be calculated using standard methods (Wolter 2007) and can be easily computed using the \CRANpkg{survey} package. Let \(V^{\texttt{HT}}_{it}\) denote the design-based variance of \(\mbox{logit}(p_{it}^{\texttt{HT}})\), obtained from the design-based variance of \(p_{it}^{\texttt{HT}}\) via linearization (the delta method). We take the logit transformed direct estimates as input data and estimate the true prevalence with the random effects model,
\begin{align}
\label{eq:smooth-direct}
\mbox{logit}(\hat p_{it}^{\texttt{HT}}) | \lambda_{it} &\sim \mbox{Normal}(\lambda_{it}, V^{\texttt{HT}}_{it}),\\
\label{eq:smooth-direct2}
\lambda_{it} &= x_{it}^\intercal\beta + \alpha_t + \epsilon_t + S_i + e_i + \delta_{it}.
\end{align}
In this model, which is a space-time smoothing extension of the Fay and Herriot (1979) model, \(\mbox{expit}(\lambda_{it})\) is the true prevalence we aim to estimate, and \(x_{it}\) are area-level covariates. The rest of the terms are normally distributed random effects including structured time trends \(\alpha_t\), unstructured, independent and identically distributed (iid), temporal terms \(\epsilon_t\), structured spatial trends \(S_i\), unstructured spatial terms \(e_i\), and space-time interaction terms \(\delta_{it}\). The terms \(e_i+S_i\) are implemented via the BYM2 parameterization (Riebler et al. 2016), a reparameterization of the classical BYM model (Besag, York, and Mollié 1991) that combines iid error terms with intrinsic conditional autoregressive (ICAR) random effects. Several different temporal models are implemented in \CRANpkg{SUMMER} for the structured temporal trends and space-time interaction effects, including random walks of order 1 and 2, and autoregressive models (Håvard Rue and Held 2005) with additional linear trends. The interaction term \(\delta_{it}\) can be one of the type I to IV interactions of the chosen temporal model and the ICAR model in space, as described in Knorr-Held (2000).
In order for the model to be identifiable, we impose sum-to-zero constraints on each group of random effects. More details on the prior choices are provided in the supplementary materials.

\hypertarget{estimating-mortality-rates-using-area-level-models}{%
\subsection{Estimating mortality rates using area-level models}\label{estimating-mortality-rates-using-area-level-models}}

For composite indicators such as mortality rates, the direct estimates require additional modeling. Here we focus on the estimation of the U5MR, one of the most critical and widely available population health indicator. The methodology and the functions in \CRANpkg{SUMMER} are readily applicable to mortality rates of other age groups as well, but we note that modeling mortality beyond age 5 is usually more challenging in practice because death becomes rarer, and survey data alone are not sufficient for reliable inference.

The \CRANpkg{SUMMER} package implements the discrete hazards model described in Mercer et al. (2015). We use discrete time survival analysis to estimate age-specific monthly probabilities of dying in user-defined age groups. We assume constant hazards within the age bands. The default choice uses the monthly age bands
\[[0, 1), [1, 12), [12, 24), [24, 36), [36, 48), [48, 60)\]
for U5MR and they can be easily specified by the user. The U5MR for area \(i\) and time \(t\) can be calculated as,
\begin{equation}
\hat p_{it}^{\texttt{HT}} = {}_{60}{\hat q}^{it}_{0} = 1 - \prod_{a = 1}^6 \left( 1 -  {}_{n_a}{\hat q}^{it}_{x_a}\right),
\end{equation}
where \(x_a\) and \(n_a\) are the start and end of the \(a\)-th age group, and \({}_{n_a}{q}^{it}_{x_a}\) is the probability of death in age group \([x_a, x_a + n_a)\) in area \(i\) and time \(t\), with \({}_{n_a}{\hat q}^{it}_{x_a}\) the estimate of this quantity. This calculation follows the synthetic cohort life table approach in which mortality probabilities for each age segments based on real cohort mortality experience are combined. This allows the full use of the most recent data, which is especially useful when survey data are sparse and is the default approach that The DHS Program (2020) uses.

The constant one-month hazards in each age band can be estimated by a weighted logistic regression model (Binder 1983):
\begin{equation}
\mbox{logit}\left(
{}_{1}{q}^{it}_{m}
\right) = \beta_{a[m]}^{it},
\end{equation}
where \(a[m]\) is the age band indicator for the \(m\)-th month, i.e.~
\begin{equation}
a[m] = \left\{
\begin{array}{ll}
1 & \mbox{if }m=0,\\
2 & \mbox{if }m=1,\dots,11,\\
3 & \mbox{if }m=12,\dots,23,\\
4 & \mbox{if }m=24,\dots,35,\\
5 & \mbox{if }m=36,\dots,47,\\
6 & \mbox{if }m=48,\dots,59.
\end{array}
\right.
\label{eq:am}
\end{equation}
The design-based variance of \(\mbox{logit}(\hat p_{it}^{\texttt{HT}})\) may then be estimated using the delta method or resampling methods such as the jackknife (J. Pedersen and Liu 2012). The smoothing of the direct estimates can then proceed using the model described in equations (\ref{eq:smooth-direct}) -- (\ref{eq:smooth-direct2}). When multiple surveys exist, one may choose to either model the survey-specific effects as fixed or random (Mercer et al. 2015) or first aggregate the direct estimates from multiple surveys to obtain a ``meta-analysis'' estimate in each area and time period (Li et al. 2019), i.e., at each time \(t\), we combine the \(K_t\) available direct estimates from multiple surveys to form the estimate\\
\[
  \hat p_{it}^{\texttt{meta}} = \mbox{expit}\Big(\sum_{k = 1}^{K_t} \frac{(\hat V_{it, k}^{\texttt{HT}})^{-1}}{\sum_{k'=1}^{K_t} (\hat V_{it, k'}^{\texttt{HT}})^{-1}} \mbox{logit}(\hat p_{it}^{\texttt{HT}}) \Big),
\]
and the associated design-based variance on the logit scale is \(\Big(\sum_{k'=1}^{K_t} (\hat V_{it, k'}^{\texttt{HT}})^{-1}\Big)^{-1}\).
To mitigate the sparsity of available data in each year, Li et al. (2019) also considers a temporal model defined at the yearly level while the direct estimates are calculated at multi-year periods. All these variations can be fit using the \CRANpkg{SUMMER} package.

\hypertarget{sec:model-cluster}{%
\subsection{Estimating mortality rates with cluster-level models}\label{sec:model-cluster}}

The space-time Fay-Herriot estimates are useful when there are enough observations at the spatial and temporal unit of the analysis. When the target of inference is at finer resolution, e.g., on a yearly time scale with admin-2 areas and surveys stratified at admin-1 levels, the direct estimates may contain many \(0\)s or \(1\)s and the design-based variance cannot be calculated reliably. In this case, we can consider unit-level models where the individual survey responses are modeled. In the rest of this section, we describe a model for the cluster-level risk, where we account for the additional within-cluster variation by allowing overdispersion in the likelihood. More detailed comparisons of this modeling choice was examined in (T. Q. Dong and Wakefield 2021). In a two-stage cluster design, the clusters are referred to as primary sampling units (PSUs) and the households are referred to as secondary sampling units (SSUs). Thus we refer to such models as \texttt{cluster-level\ model} to avoid confusion. We describe the model for the mortality estimation problem below, while the same formulation applies to the case of any generic binary indicators as well.

In the most general setting, we consider multiple surveys over time, indexed by \(k\). The sampling frame that was used for survey \(k\), will be denoted by \(r[k]\). We assume a discrete hazards model as before. We consider a beta-binomial model for the probability (hazard) of death from month \(m\) to \(m+1\) in survey \(k\) and at cluster \(c\) in year \(t\). This model allows for overdispersion relative to the binomial model. Assuming constant hazards within age bands, we assume the number of deaths occurring within age band \(a[m]\), in cluster \(c\), time \(t\), and survey \(k\) follow the beta-binomial distribution,
\begin{equation}\label{eq:BB1-age}
Y_{a[m],k,c,t} ~|~  p_{a[m],k,c,t} \sim \mbox{BetaBinomial}\left(~ n_{a[m],k,c,t}~,~ p_{m,k,c,t}~,d~
\right),
\end{equation}
where \(p_{m,k,c,t}\) is the monthly hazard at \(m\)-th month of age, in cluster \(c\), time \(t\), and survey \(k\) and \(d\) is the overdispersion parameter.
The latent logistic model we use is,
\begin{align}
p_{m,k,c,t} =& \mbox{expit}( \alpha_{m,c,k,t} +   \epsilon_t + b_k),\\ \nonumber
\alpha_{m,k,c,t} =&
\beta_{a[m],r[k],t}I(s_c \in \mbox{ rural }) +
\gamma_{a[m],r[k],t} I(s_c \in \mbox{ urban }) \\
&\;+
S_{i[s_c]} + e_{i[s_c]}  +\delta_{i[s_c],t} + \mbox{BIAS}_{k,t}.
\label{eq:pred}
\end{align}

This form consists of a collection of terms that are used for prediction and a number that are not, as we now describe. We include a survey fixed effect \(b_k\) with the constraint \(\sum_{k} b_k \mathbf{1}_{r[k] = r} = 0\) for each sampling frame \(r\), so that the main temporal trends are identifiable for each sampling frame. The \(b_k\) terms are not included in the prediction, i.e., they are set to zero. The \(\epsilon_t\) are unstructured temporal effects that allow for perturbations over time. It is a contextual choice whether they are used in predictions. We include terms in (\ref{eq:pred}) that are analogous to those in equations (\ref{eq:smooth-direct})--(\ref{eq:smooth-direct2}), in particular the spatial main effects \(S_i\) and \(e_i\) and the space-time interactions \(\delta_{it}\).

For the temporal main effects \(\beta_{a[m],r[k],t}\) and \(\gamma_{a[m],r[k],t}\), we have stratum-specific distinct trends for each age group \(a[m]\) in surveys from each sampling frame. We include separate urban and rural temporal terms to acknowledge the sampling design, often urban clusters are oversampled and have different risk to rural clusters, and so it is important to acknowledge this aspect in the model (Paige et al. 2020). The urban-rural stratification effects may also be parameterized as time-invariant fixed effects, i.e., restricting \(\beta_{a[m],r[k],t} = \gamma_{a[m],r[k],t} + \Delta_{a[m],r[k]}\). For a detailed discussion of the parameterization of stratification effects, we refer readers to Wu et al. (2021). In addition, it is usually reasonable to assume shared temporal trends up to a constant shift across some age groups. For example, we may let
\[
    \beta_{a[m],r[k],t} = \beta_{a[m],r[k], 0} + \beta^\star_{a^\star[m], r[k], t}
\]
where \(\beta^\star_{a^\star[m], r[k], t}\) is a collection of temporal random effects with sum-to-zero constraint \(\sum_{t} \beta^\star_{a^\star[m], r[k], t} = 0\), and \(a^\star[m]\) is a reduced set of age bands. The default choice for U5MR in the package is
\begin{equation}
a^\star[m] = \left\{
\begin{array}{ll}
1 & \mbox{if }m=0,\\
2 & \mbox{if }m=1,\dots,11, \\
3 & \mbox{if }m=12,\dots,59.
\end{array}
\right.
\label{eq:astar}
\end{equation}
That is, we assume the temporal trends for logit hazards in the last four age groups are parallel and only differ by the intercept term \(\beta_{a[m],r[k], 0}\).

In situations where biases are known for particular surveys and/or years, we can adjust for bias following Wakefield et al. (2019) by including the bias ratio term, \(\mbox{BIAS}_{k,t} = \mbox{U5MR}^\star_{t} / \widehat{\mbox{U5MR}}_{k, t}\), where \(\mbox{U5MR}^\star_{t}\) is the expected U5MR in year \(t\) and \(\widehat{\mbox{U5MR}}_{k, t}\) is the biased version.
This approach has been used to adjust for mothers who have died from AIDS (Walker, Hill, and Zhao 2012); such mothers cannot be surveyed, and their children are more likely to have died, so the missingness is informative.

The predicted U5MRs in urban and rural regions of area \(i\) and at time \(t\) according to sampling frame \(r\) are,
\begin{eqnarray}\label{eq:U5MR}
\mbox{U5MR}_{i,t,U,r} &=& 
1- \prod_{a=1}^6
\left[ \frac{1}{1+\exp(
\beta_{a,r,t}  + S_{i} + e_{i}  +\delta_{i,t}) }\right]^{z[a]}\\
 \mbox{U5MR}_{i,t,R,r} &=& 
1- \prod_{a=1}^6
\left[ \frac{1}{1+\exp(
\gamma_{a,r,t} + S_{i} + e_{i}  +\delta_{i,t}) }\right]^{z[a]},
\end{eqnarray}
where \(z[a]=1,11,12,12,12,12\), for the default choice of age bands.
The aggregate risk in area \(i\) and in year \(t\) according to sampling frame \(r\) is
\begin{equation}\label{eq:agg2}
p_{itr} = q_{itr}\times \mbox{U5MR}_{i,t,U,r} + (1-q_{itr})\times  \mbox{U5MR}_{i,t,R,r},
\end{equation}
where \(q_{itr}\) and \(1-q_{itr}\) are the proportions of the under-5 population in area \(i\) that are urban and rural in year \(t\) according to the classification of sampling frame \(r\). The final aggregation over different sampling frames can be done using meta-analysis combination, so that,
\[
\widehat{\mbox{U5MR}}_{it} = \mbox{expit}\left(\sum_r w_{itr} \times \mbox{logit}(p_{itr})\right),
\]
where \(w_{itr}=U_{itr}^{-1}/\sum_{r'} U_{itr'}^{-1}\) is the scaled inverse of \(U_{itr}\), which is the posterior variance of \(\mbox{logit}(\widehat{\mbox{U5MR}}_{it}^{(r)})\). Beyond point estimates, we obtain the full posterior of \(\mbox{U5MR}_{it}\), and various summaries can be reported or mapped. The estimate constructed for U5MR is not relevant to any child, because that child would have to experience the hazards for each age group simultaneously in time period \(t\), rather than moving through age groups over multiple time periods. Nevertheless, the resultant U5MR is a useful summary and the conventional measure that is used to inform on child mortality.

\hypertarget{sec:summer-overview}{%
\section{Overview of SUMMER}\label{sec:summer-overview}}

The \CRANpkg{SUMMER} package provides a collection of functions for SAE with complex survey data. The package can be installed in R directly by

\begin{verbatim}
install.packages("SUMMER")
\end{verbatim}

The \CRANpkg{SUMMER} package requires the \pkg{INLA} package (H. Rue, Martino, and Chopin 2009) to be installed. All analysis in this package are conducted with \CRANpkg{SUMMER} package version \(1.4.0\) and \pkg{INLA} version \(24.03.09\). \pkg{INLA} can be installed with

\begin{verbatim}
  install.packages("INLA", repos=c(getOption("repos"), 
                    INLA="https://inla.r-inla-download.org/R/stable"), dep=TRUE)
\end{verbatim}

The \CRANpkg{SUMMER} package implements a variety of space-time smoothing models using survey data. There are three main functions to implement these models, discussed below and in the three examples in the following sections.

\begin{itemize}
\tightlist
\item
  \texttt{smoothSurvey()} produces direct and Fay-Herriot estimates for a generic binary indicator from raw survey data.
\item
  \texttt{smoothDirect()} takes direct estimates as input and produces the Fay-Herriot estimates for mortality estimation discussed in Li et al. (2019).\\
\item
  \texttt{smoothCluster()} performs cluster-level smoothing using the Beta-Binomial model discussed in the previous section, and with more details discussed in Wu et al. (2021) and Fuglstad, Li, and Wakefield (2021).
\end{itemize}

We note that \texttt{smoothDirect()} and \texttt{smoothCluster()} includes many more features in modeling composite indicators such as child mortality. In comparison, \texttt{smoothSurvey()} can only model generic binary indicators, but it has a simpler interface and workflow, which is appealing for broader communities of practitioners.

\begin{figure}[!ht]
\includegraphics[width=1\linewidth,]{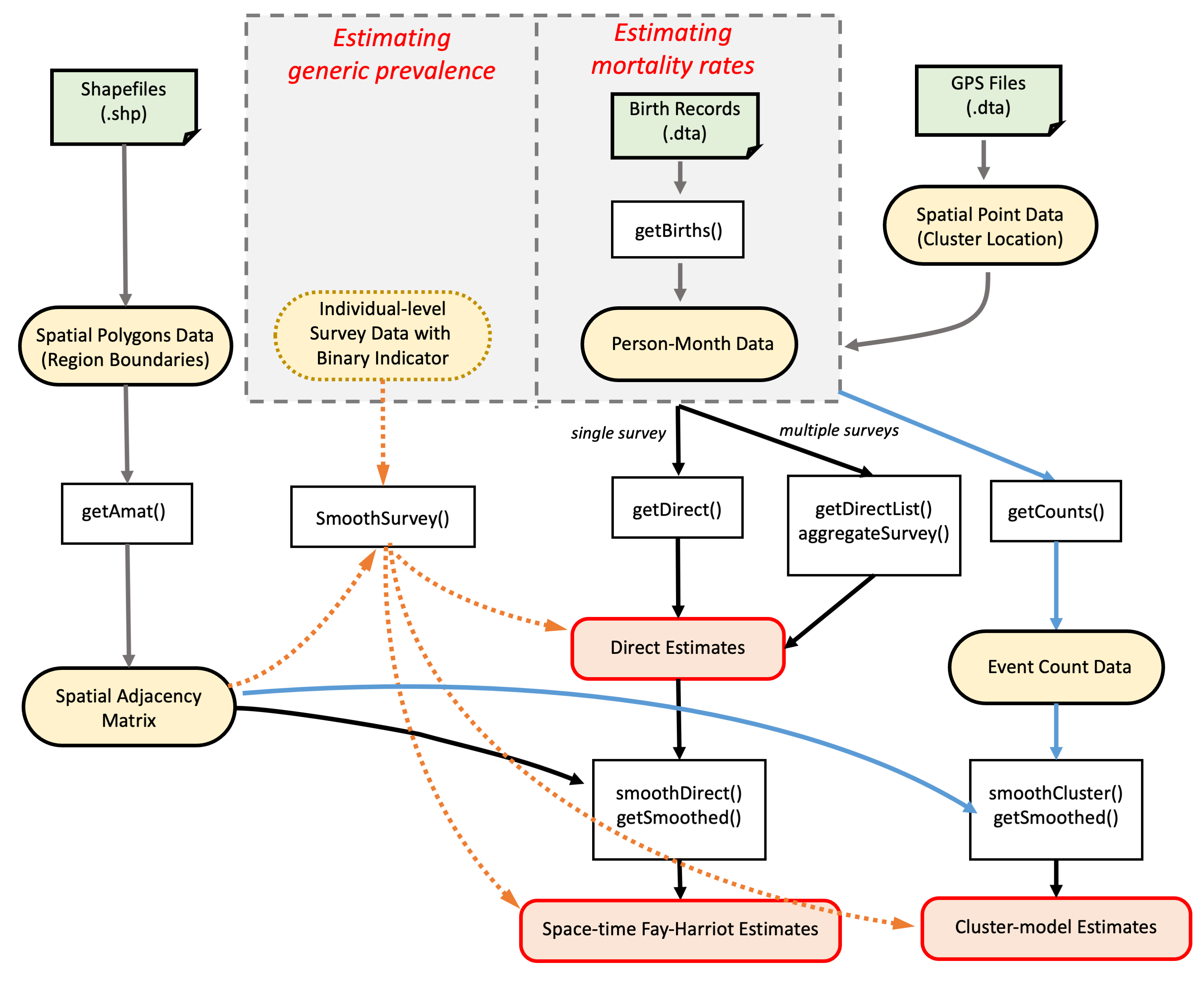} \caption{Workflow of the SUMMER package. Rounded blocks represent data types and rectangular blocks represent functions in the SUMMER package. Output estimates are highlighted in the boxes with red borders. The dotted yellow arrows represent the workflow using \texttt{smoothSurvey()} to estimate the prevalence of a generic binary indicator. The black solid arrows represent the workflow using \texttt{smoothDirect()} to perform area-level smoothing of mortality rates. The blue solid arrows represent the workflow using \texttt{smoothCluster()} to perform cluster-level smoothing of mortality rates.}\label{fig:workflow}
\end{figure}

The main source of data required for these methods are the survey data and the corresponding spatial adjacency matrix, which can be derived from spatial polygons data describing region boundaries. For cluster-level modeling, we also need to know which region each cluster belongs to. In the context of modeling DHS data, the survey data and cluster locations are usually recorded in separate files. Figure \ref{fig:workflow} shows schematically the workflow of data processing and smoothing for generic binary indicators and mortality estimates using the \CRANpkg{SUMMER} package. In this paper, our workflow starts with the birth records and GPS files in .dta files as an example. Such files can be directly downloaded from the DHS data portal. It is also straightforward to load data in other formats and supply the R objects into the functions. The entire pipeline of analysis can be carried out using functions in \CRANpkg{SUMMER}. The analysis of the main paper can be reproduced without registering for data access. We include a more extensive analysis of DHS data in the supplementary materials, which requires registration with DHS program for data access.

Before demonstrating the utilities of these functions in the following examples, we first load the packages for the analysis, data processing and visualization. For the analysis presented in this paper, we use the \CRANpkg{ggplot2} package (Wickham 2016) and \CRANpkg{patchwork} package (T. L. Pedersen 2019) to make further customize the visualization produced by \CRANpkg{SUMMER}.

\begin{verbatim}
library(SUMMER)
library(ggplot2)
library(patchwork)
\end{verbatim}

\hypertarget{example-1-prevalence-estimation-for-a-binary-indicator}{%
\section{Example 1: prevalence estimation for a binary indicator}\label{example-1-prevalence-estimation-for-a-binary-indicator}}

We start by considering the simplest scenario of estimating the prevalence of a binary indicator using a dataset from the Behavioral Risk Factor Surveillance System (BRFSS) survey. BRFSS is an annual telephone health survey conducted by the Centers for Disease Control and Prevention (CDC) that tracks health conditions and risk behaviors in the United States and its territories since 1984. The BRFSS sampling scheme is complex with high variability in the sampling weights. In this example, we estimate the prevalence of Type II diabetes in health reporting areas (HRAs) in the King County of Washington using BRFSS data. We will compare the direct estimates and the Fay-Herriot estimates.

The \texttt{BRFSS} dataset in \CRANpkg{SUMMER} contains the full BRFSS dataset with \(16,283\) observations. The \texttt{diab2} variable is the binary indicator of Type II diabetes, \texttt{strata} is the strata indicator and \texttt{rwt\_llcp} is the final design weight. For the purpose of this analysis, we first remove records with missing HRA code or diabetes status from this dataset.

\begin{verbatim}
data(BRFSS)
data <- subset(BRFSS, !is.na(diab2) & !is.na(hracode))
\end{verbatim}

The \texttt{KingCounty} dataset in \CRANpkg{SUMMER} contains the map of the HRAs in the King County. We first extract the spatial adjacency matrix for the HRAs using the \texttt{getAmat()} function.

\begin{verbatim}
data(KingCounty)
KingGraph <- getAmat(KingCounty, KingCounty$HRA2010v2_)
\end{verbatim}

We then use the \texttt{smoothSurvey()} function to obtain both the direct and Fay-Herriot estimates by HRA. The function requires specifications of the variables that determines the survey design, including sampling weights (\texttt{weightVar}), strata indicator (\texttt{strataVar}), and cluster identifiers (\texttt{clusterVar}). In this dataset, there are no clusters so we use the formula \texttt{\textasciitilde{}1} in this situation (Lumley 2004). We also need to specify region indicators (\texttt{regionVar}) in the data frame that match the column and row names of the spatial adjacency matrix.

\begin{verbatim}
fit.BRFSS <- smoothSurvey(data = data, Amat = KingGraph, 
                          response.type = "binary", responseVar = "diab2", 
                          strataVar="strata", weightVar="rwt_llcp", 
                          regionVar="hracode", clusterVar = "~1")
head(fit.BRFSS$direct, n = 3)
\end{verbatim}

\begin{verbatim}
#>         region direct.est direct.var direct.logit.est direct.logit.var direct.logit.prec
#> 1 Auburn-North       0.10    0.00046             -2.2            0.053              18.8
#> 2 Auburn-South       0.23    0.00240             -1.2            0.075              13.3
#> 3      Ballard       0.07    0.00050             -2.6            0.115               8.7
\end{verbatim}

\begin{verbatim}
head(fit.BRFSS$smooth, n = 3)
\end{verbatim}

\begin{verbatim}
#>         region  mean     var median lower upper logit.mean logit.var logit.median
#> 1 Auburn-North 0.102 0.00026  0.101 0.074 0.137       -2.2     0.030         -2.2
#> 2 Auburn-South 0.160 0.00093  0.157 0.108 0.228       -1.7     0.051         -1.7
#> 3      Ballard 0.059 0.00018  0.057 0.037 0.089       -2.8     0.057         -2.8
#>   logit.lower logit.upper
#> 1        -2.5        -2.5
#> 2        -2.1        -2.1
#> 3        -3.3        -3.3
\end{verbatim}

\begin{figure}[!ht]
\includegraphics[width=1\linewidth,]{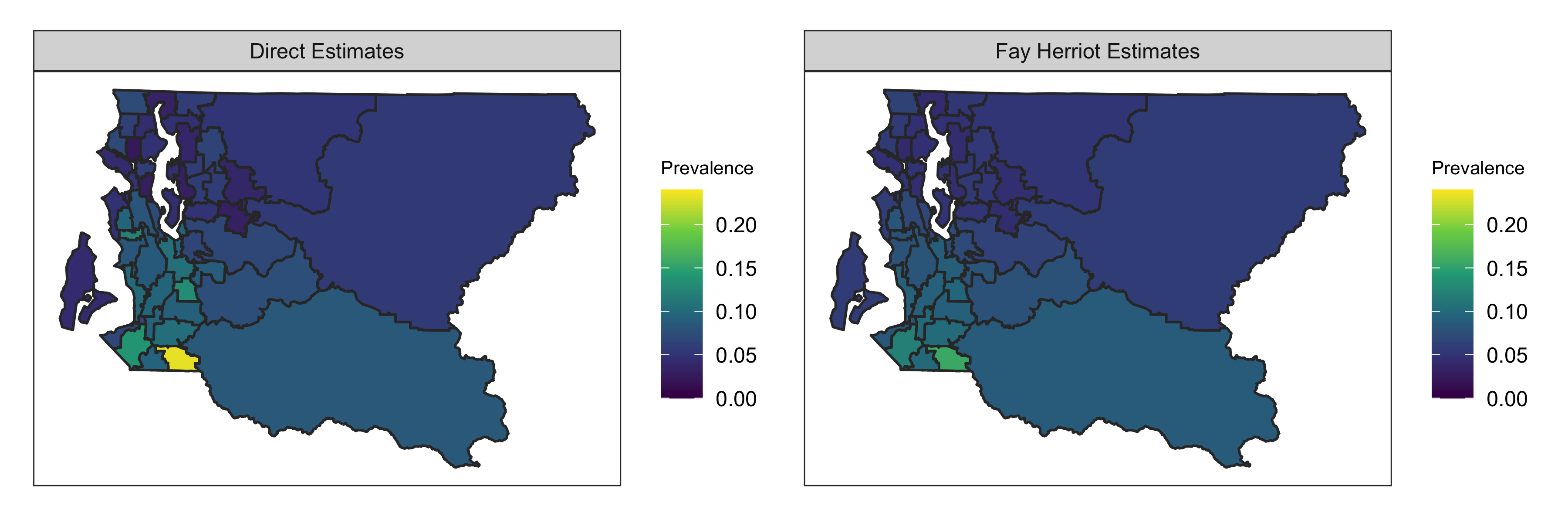} \caption{Direct and Fay-Herriot estimates of the prevalence of Type II diabetes in King county HRAs.}\label{fig:brfss-4}
\end{figure}

The fitted object is of class \texttt{SUMMERmodel.svy}, and the direct (\texttt{\$direct}) and Fay-Herriot estimates (\texttt{\$smooth}) are saved as data frames in the fitted objects. Notice that since the analysis is performed on the logit of the prevalence, estimates on both the logit and the probability scales are returned in the output. We use the \texttt{mapPlot()} function in \CRANpkg{SUMMER} to show the estimates on a map. In essence, the \texttt{mapPlot()} function takes a data frame, a \texttt{SpatialPolygonsDataFrame} object, the column names indexing regions in both the data frame and the polygons, and returns a \texttt{ggplot} object. Additional function arguments are available to more easily customize the visualizations. Figure \ref{fig:brfss-4} compares the point estimates on the map and the effect of spatial smoothing can be easily seen.

\begin{verbatim}
g1 <- mapPlot(fit.BRFSS$direct, geo = KingCounty, 
              by.data = "region", by.geo = "HRA2010v2_", 
              variables = "direct.est", label = "Direct Estimates", 
              legend.label = "Prevalence", ylim = c(0, 0.24))
g2 <- mapPlot(fit.BRFSS$smooth, geo = KingCounty, 
              by.data = "region", by.geo = "HRA2010v2_", 
              variables = "median", label = "Fay Herriot Estimates", 
              legend.label = "Prevalence", ylim = c(0, 0.24))
g1 + g2
\end{verbatim}

Similar analysis can be implemented for Gaussian observations, and can include temporal smoothing or covariates in the smoothing model. The hyperpriors can also be further customized. For more details and examples, we refer the readers to the package documentation under the \texttt{smoothSurvey()} section.

\hypertarget{example-2-area-level-model-of-nmr-and-u5mr-using-simulated-data}{%
\section{Example 2: area-level model of NMR and U5MR using simulated data}\label{example-2-area-level-model-of-nmr-and-u5mr-using-simulated-data}}

In the second example, we consider estimating mortality rates using multiple surveys. We use the NMR and U5MR as two examples, but the implementation in the \CRANpkg{SUMMER} package allows straightforward extensions to other age groups. We use a simulated survey dataset in this example. A more detailed case study using cluster-level model is provided in the supplementary materials.

We load the \texttt{DemoData} dataset from the \CRANpkg{SUMMER} package. The \texttt{DemoData} is a list that contains full birth history data from simulated surveys with stratified cluster sampling design, similar to most of the DHS surveys. It has been pre-processed into the person-month format, where for each list entry, each row represents one person-month record. Each record contains columns for the cluster ID (\texttt{clustid}), household ID (\texttt{id}), strata membership (\texttt{strata}) and survey weights (\texttt{weights}). The region and time period associated with each person-month record has also been pre-computed. The age variable in this data frame are in the form of \(a_1\)-\(a_2\), i.e., \(1\)-\(11\) corresponds to age group with 1 to 11 completed months, whereas age groups with only one month are stored using a single number representation, e.g., age group \(0\). This is also the data structure in the output of the \code{getBirths} function in the \CRANpkg{SUMMER} package. In all the analysis of this paper, we use the default age bands as described before. If a different set of age band is desired, they can be specified by the \texttt{month.cut} argument in the \code{getBirths} function.

\begin{verbatim}
data(DemoData)
head(DemoData[[1]])
\end{verbatim}

\begin{verbatim}
#>   clustid id  region  time  age weights        strata died
#> 1       1  1 eastern 00-04    0     1.1 eastern.rural    0
#> 2       1  1 eastern 00-04 1-11     1.1 eastern.rural    0
#> 3       1  1 eastern 00-04 1-11     1.1 eastern.rural    0
#> 4       1  1 eastern 00-04 1-11     1.1 eastern.rural    0
#> 5       1  1 eastern 00-04 1-11     1.1 eastern.rural    0
#> 6       1  1 eastern 00-04 1-11     1.1 eastern.rural    0
\end{verbatim}

In order to compute NMR, we create a new list of surveys with only deaths within age group \(0\).

\begin{verbatim}
DemoDataNMR <- DemoData
for(i in 1:length(DemoData)){
    DemoDataNMR[[i]] <- subset(DemoData[[i]], age == "0")
}
\end{verbatim}

We now turn to the estimation of NMR and U5MR using four simulated surveys in \texttt{DemoData}. For multiple surveys, we combine the person-month records into a list and use the \texttt{getDirectList()} function to obtain the survey specific direct estimates. When there are no deaths in a given area and time period, or when more than half of the age groups do not exist in the person-month data, the direct estimates cannot be reliably computed and are set to NA. When only a small fraction of the age groups are not observed, they will be combined with the previous age groups when fitting the discrete hazard model.

\begin{verbatim}
periods <- c("85-89", "90-94", "95-99", "00-04", "05-09", "10-14")
directNMR <- getDirectList(births = DemoDataNMR, years = periods, 
                           regionVar = "region", timeVar = "time", 
                           clusterVar = "~clustid + id", ageVar = "age", 
                           weightsVar = "weights")
directU5 <- getDirectList(births = DemoData, years = periods, 
                          regionVar = "region", timeVar = "time", 
                          clusterVar = "~clustid + id", ageVar = "age", 
                          weightsVar = "weights")
\end{verbatim}

The direct estimates from multiple surveys can be combined to produce a ``meta-analysis'' estimator using the \texttt{aggregateSurvey()} function.

\begin{verbatim}
directNMR.comb <- aggregateSurvey(directNMR)
directU5.comb <- aggregateSurvey(directU5)
\end{verbatim}

Once the direct estimates are calculated, we fit the space-time Fay-Herriot model in the same fashion as in the previous example. The argument \texttt{year.label} specifies the order of the years column in the direct estimates, so that it does not have to be integer valued, and can easily allow extensions to future and past time periods not in the data. We can also fit the temporal model at the yearly level even though the direct estimates are in five year periods (Li et al. 2019). In this case we need to specify the proper range of the time periods (\texttt{year\_range}) encoded by the time periods in \texttt{year.label}, and the number of years in each period (\texttt{m}). Unequal periods are not supported at this time. The \texttt{smoothDirect()} function returns a fitted object of class \texttt{SUMMERmodel}.

\begin{verbatim}
fhNMR <- smoothDirect(data = directNMR.comb, Amat = DemoMap$Amat, 
                      year.label = c(periods, "15-19"), year.range = c(1985, 2019), 
                      time.model = "rw2", type.st = 4, is.yearly = TRUE, m = 5)
fhU5 <- smoothDirect(data = directU5.comb, Amat = DemoMap$Amat, 
                      year.label = c(periods, "15-19"), year.range = c(1985, 2019), 
                      time.model = "rw2", type.st = 4, is.yearly = TRUE, m = 5)
\end{verbatim}

\begin{figure}[!ht]
\includegraphics[width=1\linewidth,]{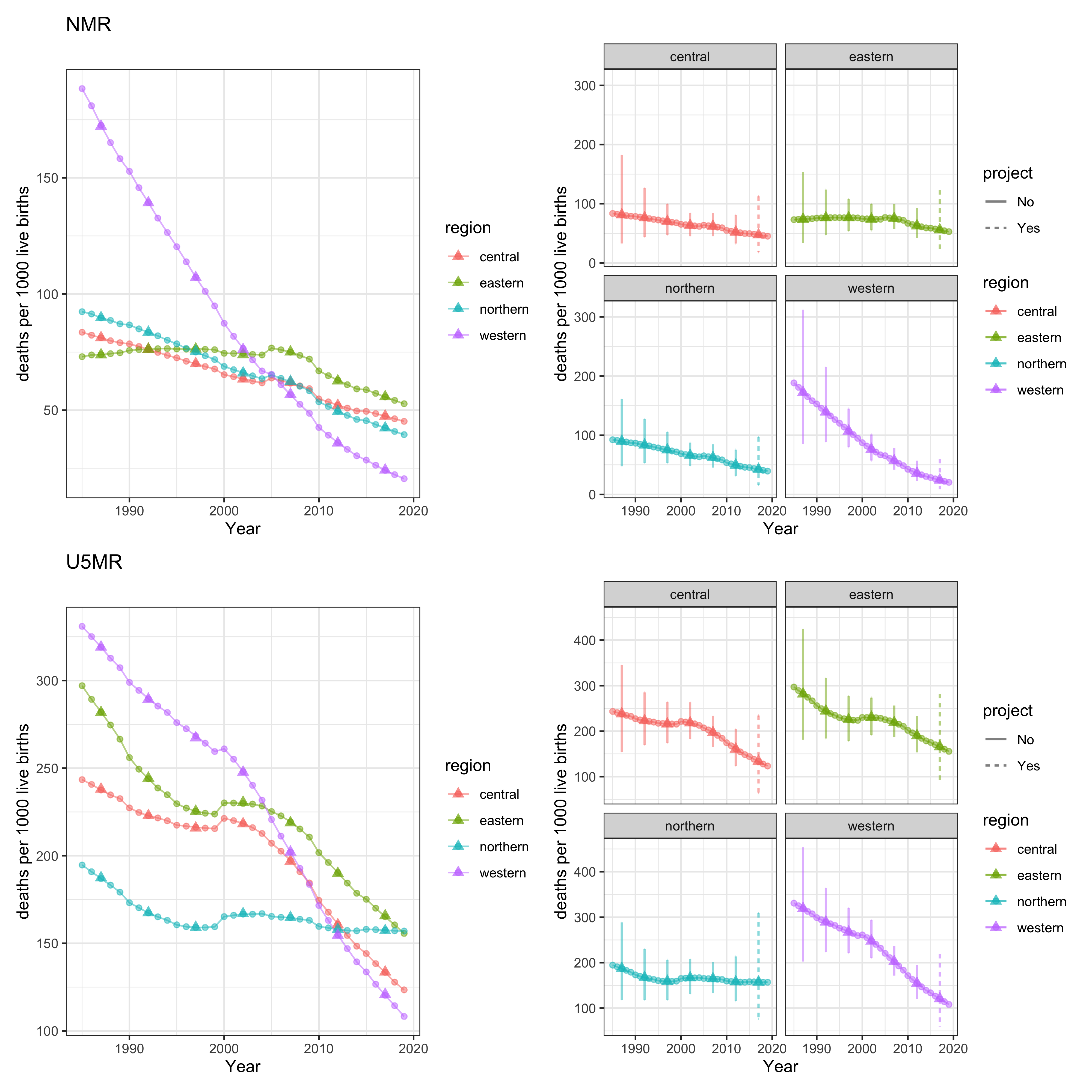} \caption{Smoothed direct estimates of NMR (top row) and U5MR (bottom row) on the yearly scale (dots) and 5-year period scale (triangles) in the simulated dataset. The vertical error bars correspond to 95\% credible interval of the 5-year estimates. The plots on the left are from the default plot function. The plots on the right shows simple customization of the default plots.}\label{fig:smooth-direct-estimate-2}
\end{figure}

The Fay-Herriot estimates can be summarized by the \texttt{getSmoothed()} function. The desired posterior credible intervals are specified by the \texttt{CI} argument. It organizes the estimates into a data frame of class \texttt{SUMMERproj}, which can be directly viewed or plotted using the \texttt{plot} method. Additional customization can be added using the syntax of \CRANpkg{ggplot2}, as shown in Figure \ref{fig:smooth-direct-estimate-2}.

\begin{verbatim}
est.NMR <- getSmoothed(fhNMR, CI = 0.95)
est.U5 <- getSmoothed(fhU5, CI = 0.95)
g3 <- plot(est.NMR, per1000 = TRUE) + ggtitle("NMR")
g4 <- plot(est.NMR, per1000 = TRUE, plot.CI=TRUE) + facet_wrap(~region)
g5 <- plot(est.U5, per1000 = TRUE)  + ggtitle("U5MR")
g6 <- plot(est.U5, per1000 = TRUE, plot.CI=TRUE) + facet_wrap(~region)
(g3 + g4) / (g5 + g6)
\end{verbatim}

\hypertarget{example-3-cluster-level-model-of-u5mr-using-malawi-dhs-data}{%
\section{Example 3: cluster-level model of U5MR using Malawi DHS data}\label{example-3-cluster-level-model-of-u5mr-using-malawi-dhs-data}}

We now consider a more realistic example of estimating U5MR at admin-2 level using the 2015--2016 Malawi DHS survey. The full dataset is available on the DHS website at \url{https://dhsprogram.com/data/available-datasets.cfm?ctryid=24}. Access to the full micro-level data requires registration with the DHS. Once access is approved, the \CRANpkg{rdhs} (Watson and Eaton 2019) package can be used to load data directly from the DHS API in R. We document the process to read the raw DHS files and process the birth records in the supplementary materials.

For reproducibility of the examples in this paper, we start with the aggregated count data from the 2015 Malawi DHS. The pre-processed count data is available in the supplementary materials. This aggregated dataset consists of the counts of deaths occurred within each age band and the total number of person-months by cluster and year. This aggregated dataset and the full data acquisition and cleaning steps to obtain this dataset are described in the supplementary materials. The processing steps involves primarily the \texttt{getBirths()} and \texttt{getCounts()} functions and some data cleaning in region names. The supplementary materials also includes workflows and results on fitting several other smoothing models on the Malawi DHS data.

Subnational spatial polygon files can usually be found on the DHS spatial data repository (The DHS Program 2020) or the GADM database of global administrative areas (Global Administrative Areas 2012). The admin-2 region polygon of Malawi is included in the \CRANpkg{SUMMER} package already and can be directly loaded.

\begin{verbatim}
data(MalawiMap)
MalawiGraph = getAmat(MalawiMap, names=MalawiMap$ADM2_EN)
\end{verbatim}

We then load the pre-processed count data and fit the cluster-level model using the \texttt{smoothCluster()} function. We consider the observations from 2007 to 2015 and project the mortality rates to 2019. To simplify results, we fit the unstratified model in this example by removing the \texttt{strata} variable from the data frame or setting it to \texttt{NA}. The supplementary materials contains additional details to fit stratified cluster-level models.

\begin{verbatim}
load("Data/DHS_counts.rda")
agg.counts$strata <- NA
head(agg.counts)
\end{verbatim}

\begin{verbatim}
#>       v001  v025 admin2 time age  v005 died total  survey cluster strata region years Y
#> 55427  696 urban Likoma 2000   0 12778    0     3 DHS2015     696     NA Likoma  2000 0
#> 55428  696 urban Likoma 2001   0 12778    0     4 DHS2015     696     NA Likoma  2001 0
#> 55429  696 urban Likoma 2002   0 12778    0     2 DHS2015     696     NA Likoma  2002 0
#> 55430  696 urban Likoma 2003   0 12778    0     4 DHS2015     696     NA Likoma  2003 0
#> 55431  696 urban Likoma 2004   0 12778    0     3 DHS2015     696     NA Likoma  2004 0
#> 55432  696 urban Likoma 2005   0 12778    0     2 DHS2015     696     NA Likoma  2005 0
\end{verbatim}

Sometimes additional information is available to adjust the estimates from the surveys. For example, in countries with high prevalence of HIV, estimates of U5MR can be biased, particularly before ART treatment became widely available. Pre-treatment, HIV positive women had a high risk of dying, and such women who had given birth were therefore less likely to appear in surveys. The children of HIV positive women are also more likely to have a higher probability of dying compared to those born to HIV negative women. Hence, we expect that the U5MR is underestimated if we do not adjust for the missing women. For the two surveys in Malawi, the calculated HIV adjustment ratios as described in Walker, Hill, and Zhao (2012) are stored in the \CRANpkg{SUMMER} as \texttt{MalawiData\$HIV.yearly}. The unstratified cluster-level model can be fitted using the \texttt{smoothCluster()} function.

\begin{verbatim}
fit.bb <- smoothCluster(data = agg.counts, Amat = MalawiGraph, 
                        family = "betabinomial", year.label = 2000:2019, 
                        time.model = "rw2", st.time.model = "ar1",
                        age.group = c("0", "1-11", "12-23", "24-35", "36-47", "48-59"),
                        age.n = c(1, 11, 12, 12, 12, 12), 
                        age.time.group = c(1, 2, 3, 3, 3, 3),
                        pc.st.slope.u = 2, pc.st.slope.alpha = 0.1,
                        bias.adj = MalawiData$HIV.yearly, 
                        bias.adj.by = c("years", "survey"),
                        survey.effect = FALSE)
\end{verbatim}

When not specified explicitly, the space-time interaction term inherits the same temporal dependency structure defined by \texttt{time.model}. We can use different models for the interaction term by specifying \texttt{st.time.model} argument. For example, in the model above, we can model the main temporal trends using random walks of order 2, and model the space-time interaction using the interaction of a temporal AR(1) process and an ICAR process in space. To allow each region to have more flexible temporal trends, we add region-specific random slopes to the interaction terms by specifying the priors \texttt{pc.st.slope.u} and \texttt{pc.st.slope.alpha}. These arguments specifies that the probability of the absolute temporal change from the shared temporal trend (on the logit scale) over the entire time period exceeding \texttt{pc.st.slope.u} is \texttt{pc.st.slope.alpha}. The \texttt{age.group}, \texttt{age.n} and \texttt{age.time.group} specify the age groups, their corresponding length (in months), and how they are grouped when modeling the temporal trends, i.e., the \(\alpha^\star[m]\) term defined before.

After we fit the model, we use the \texttt{getSmoothed()} function to obtain the posterior summaries of the prevalence by taking \texttt{nsim} draws from the posterior distribution. Since for the cluster-level models, the estimates may not be a linear combination of the random effect terms in the case of a composite indicator, the posterior summaries are obtained via posterior samples. For the cluster-level model, the \texttt{getSmoothed()} function returns a few an object of class \texttt{SUMMERprojlist}, which includes potentially multiple projections for each stratum (\texttt{\$stratified}), sampling frame (\texttt{\$overall}), and aggregated over different sampling frames (\texttt{\$final}) if applicable. In this example, since we fit an unstratified model with only one sampling frame, the estimates stored in \texttt{est.bb\$stratified} and \texttt{est.bb\$overall} are the same. In addition, by specifying \texttt{save.draws\ =\ TRUE}, the full posterior draws are stored, which can be re-used to speed up other functions that require access to posterior samples of internal model parameters.

\begin{verbatim}
est.bb <-  getSmoothed(fit.bb, nsim = 1000, save.draws = TRUE) 
\end{verbatim}

\hypertarget{visualization-and-model-checking}{%
\section{Visualization and model checking}\label{visualization-and-model-checking}}

In addition to the plots shown in the previous sections, the \CRANpkg{SUMMER} package provides a collection of visualization tools to assess model fit and uncertainty in the estimates. Assessment of uncertainty is a key step in analysis as maps of point estimates can be intoxicating, but often hide huge uncertainty, which should temper initial enthusiasm. Most of the visualization options return a \CRANpkg{ggplot2} object, which can be further customized. In this section, we use the fitted models for Malawi 2015 -- 2016 DHS as an example.

\begin{figure}[!ht]

{\centering \includegraphics[width=0.5\linewidth,]{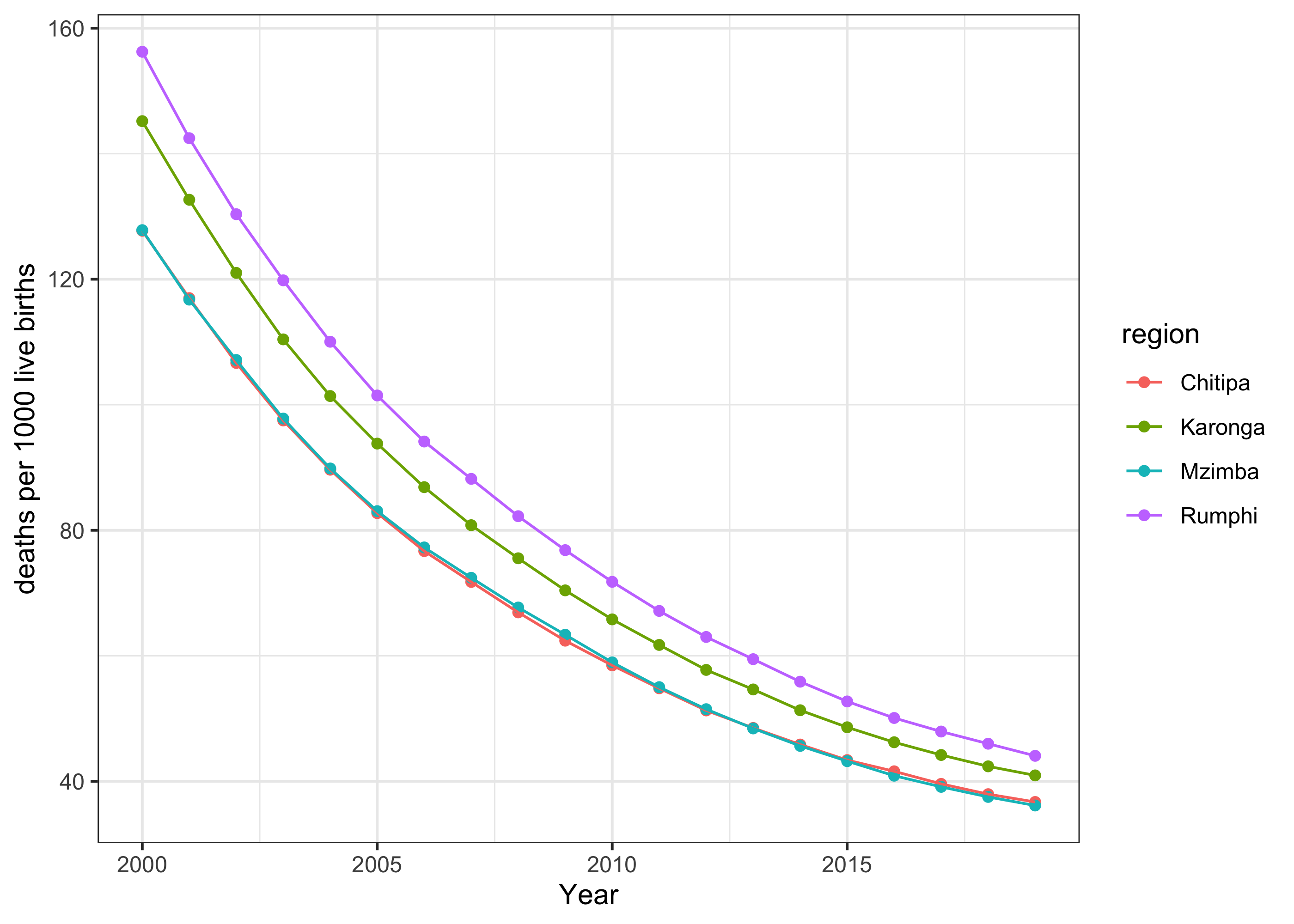} 

}

\caption{Subnational temporal trends of U5MR using the 2015--2016 DHS in Malawi in four regions.}\label{fig:vis-1}
\end{figure}

The first set of visualizations are the line plots that we have shown before. Figure \ref{fig:vis-1} shows subnational posterior median U5MR estimates over time for the five northern regions. We also scale the estimates to be deaths per \(1,000\) live births using the \texttt{per1000} argument.

\begin{verbatim}
select <- c("Chitipa", "Karonga", "Rumphi", "Mzimba")
plot(subset(est.bb$overall, region %in% select), per1000 = TRUE, year.proj = 2016)
\end{verbatim}

\begin{figure}[!ht]

{\centering \includegraphics[width=1\linewidth,]{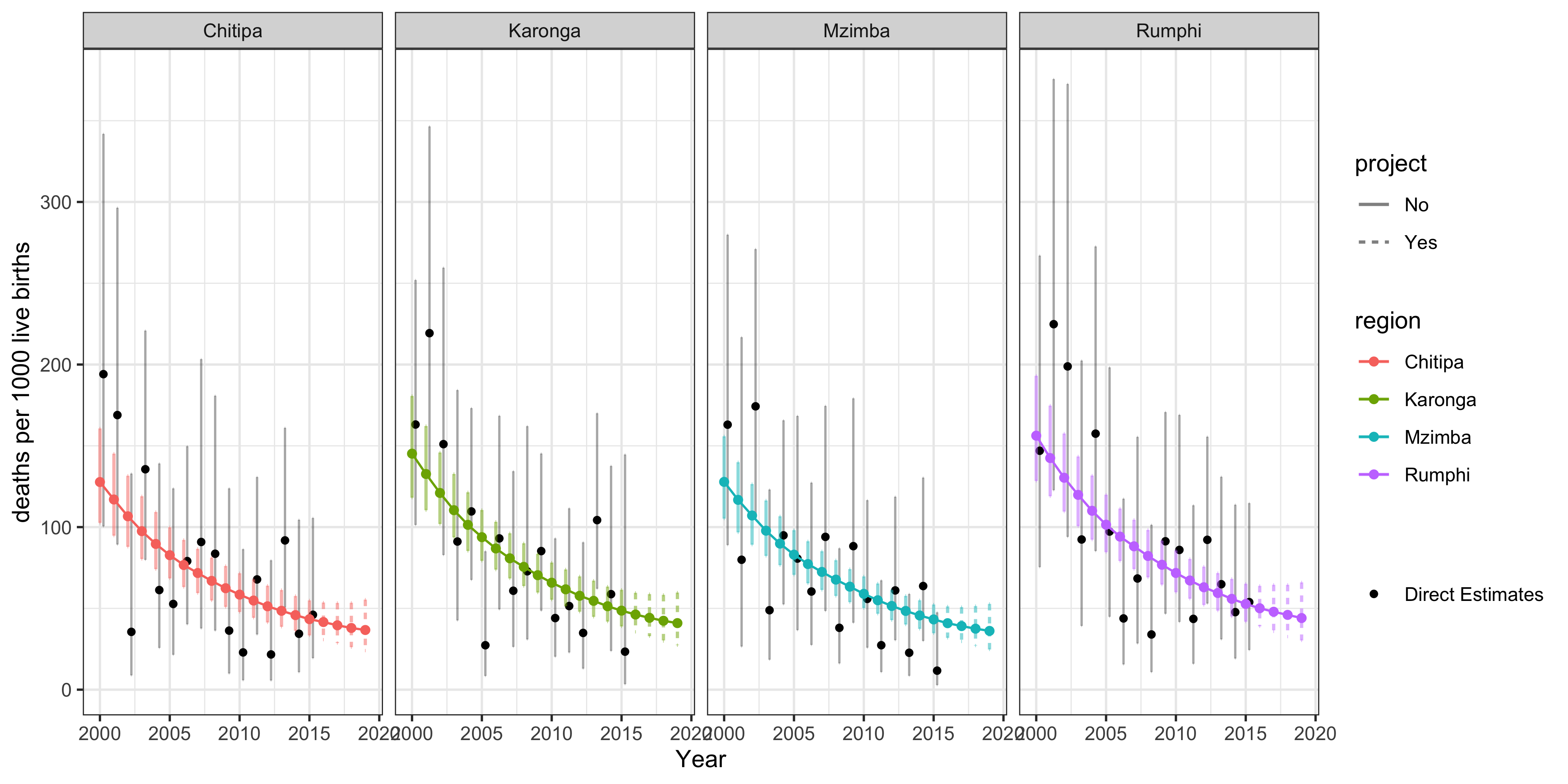} 

}

\caption{Comparing subnational temporal trends of U5MR under cluster-level model and direct estimates, using the 2015--2016 DHS in Malawi in four regions.}\label{fig:vis-1b}
\end{figure}

By default, subnational estimates do not show the intervals to avoid many overlapping vertical bars, but they can be added back with the \texttt{plot.CI} option as illustrated in previous examples. We also compare the smoothed estimates with the pre-computed direct estimates in Figure \ref{fig:vis-1b}, where the shrinkage in point estimates and the reduction in uncertainty intervals can be easily seen. The direct estimate computation are detailed in the supplementary materials.

\begin{verbatim}
load("Data/DHS_direct_hiv_adj.rda")
plot(subset(est.bb$overall, region %in% select), per1000 = TRUE, 
     year.proj = 2016, plot.CI = TRUE,
     data.add = direct.2015.hiv, label.add = "Direct Estimates", 
     option.add = list(point = "mean", lower = "lower", upper = "upper")) + 
facet_wrap(~region, ncol = 4)
\end{verbatim}

\begin{figure}[!ht]
\includegraphics[width=1\linewidth,]{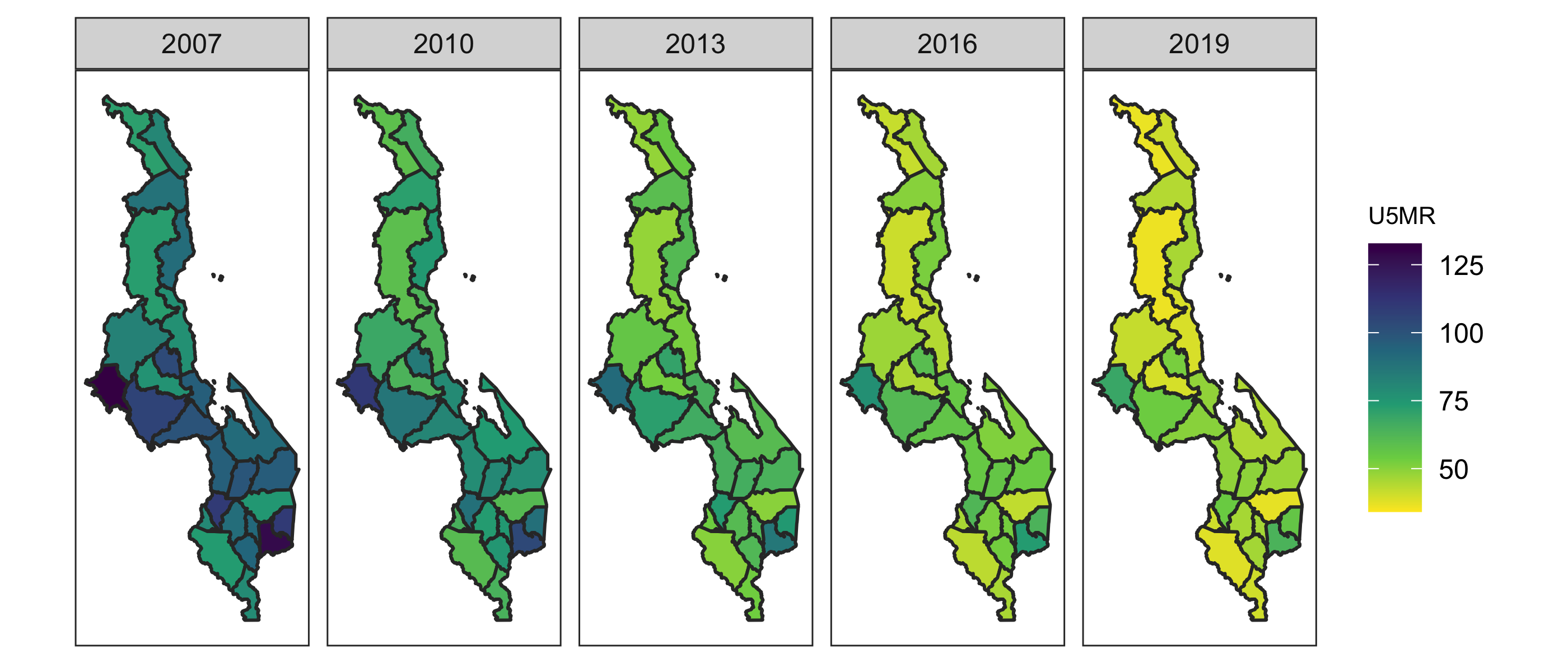} \caption{Spatial distribution of U5MR using the 2015--2016 DHS in Malawi over selected years.}\label{fig:vis-2}
\end{figure}

The \texttt{mapPlot()} function visualizes the estimates on a map. The estimates, \texttt{est.bb\$overall}, are in the long format where estimates of each year and period are stacked. This is specified with the \texttt{is.long} argument. Figure \ref{fig:vis-2} maps the changes over time. The drops in U5MR in all regions is apparent, though there is great spatial heterogeneity.

\begin{verbatim}
year.plot <- c("2007", "2010", "2013", "2016", "2019")
mapPlot(subset(est.bb$overall, years %in% year.plot), 
        geo = MalawiMap, by.data = "region", by.geo = "ADM2_EN", 
        is.long = TRUE, variables = "years", values = "median", 
        ncol = 5, direction = -1, per1000 = TRUE, legend.label = "U5MR")
\end{verbatim}

\begin{figure}[!ht]
\includegraphics[width=1\linewidth,]{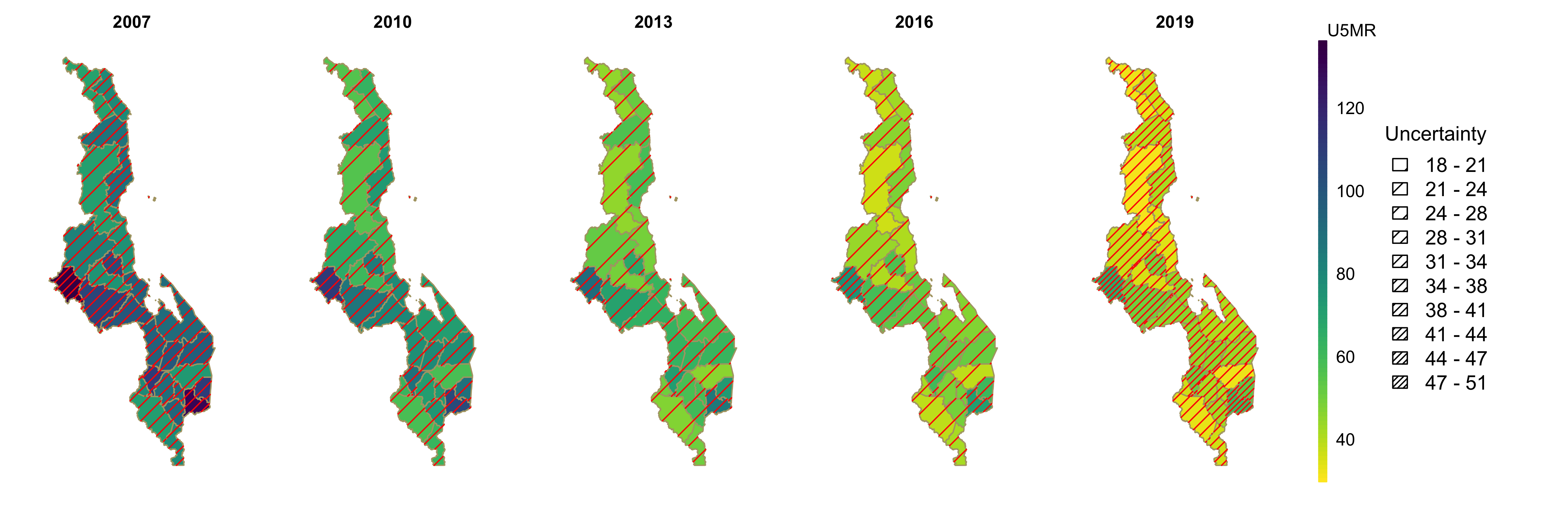} \caption{Subnational estimates of U5MR using the 2015--2016 DHS in Malawi over selected years, with hatching lines indicating the width of the 95\% credible intervals of the estimates. Denser hatching correspond to higher uncertainty. Estimates for 2019 in the last column are from the model projection and thus have higher uncertainty.}\label{fig:vis-3}
\end{figure}

The \texttt{hatchPlot()} function plots additional hatching lines on the map indicating the width of the uncertainty intervals. Denser hatching lines represent higher uncertainty. Usually estimates of the early years have higher uncertainty as shown in Figure \ref{fig:vis-3}. It also clearly shows the increase in uncertainty in the projections. We also note that both \texttt{mapPlot()} and \texttt{hatchPlot()} function can be used in broader cases as they provide a general tool to visualize rectangular data on a map.

\begin{verbatim}
hatchPlot(subset(est.bb$overall, years %in% year.plot), 
          geo = MalawiMap, by.data = "region", by.geo = "ADM2_EN", 
          is.long = TRUE, variables = "years", values = "median", 
          lower = "lower", upper = "upper", hatch = "red",
          ncol = 5, direction = -1, per1000 = TRUE, legend.label = "U5MR")
\end{verbatim}

\begin{figure}[!ht]
\includegraphics[width=1\linewidth,]{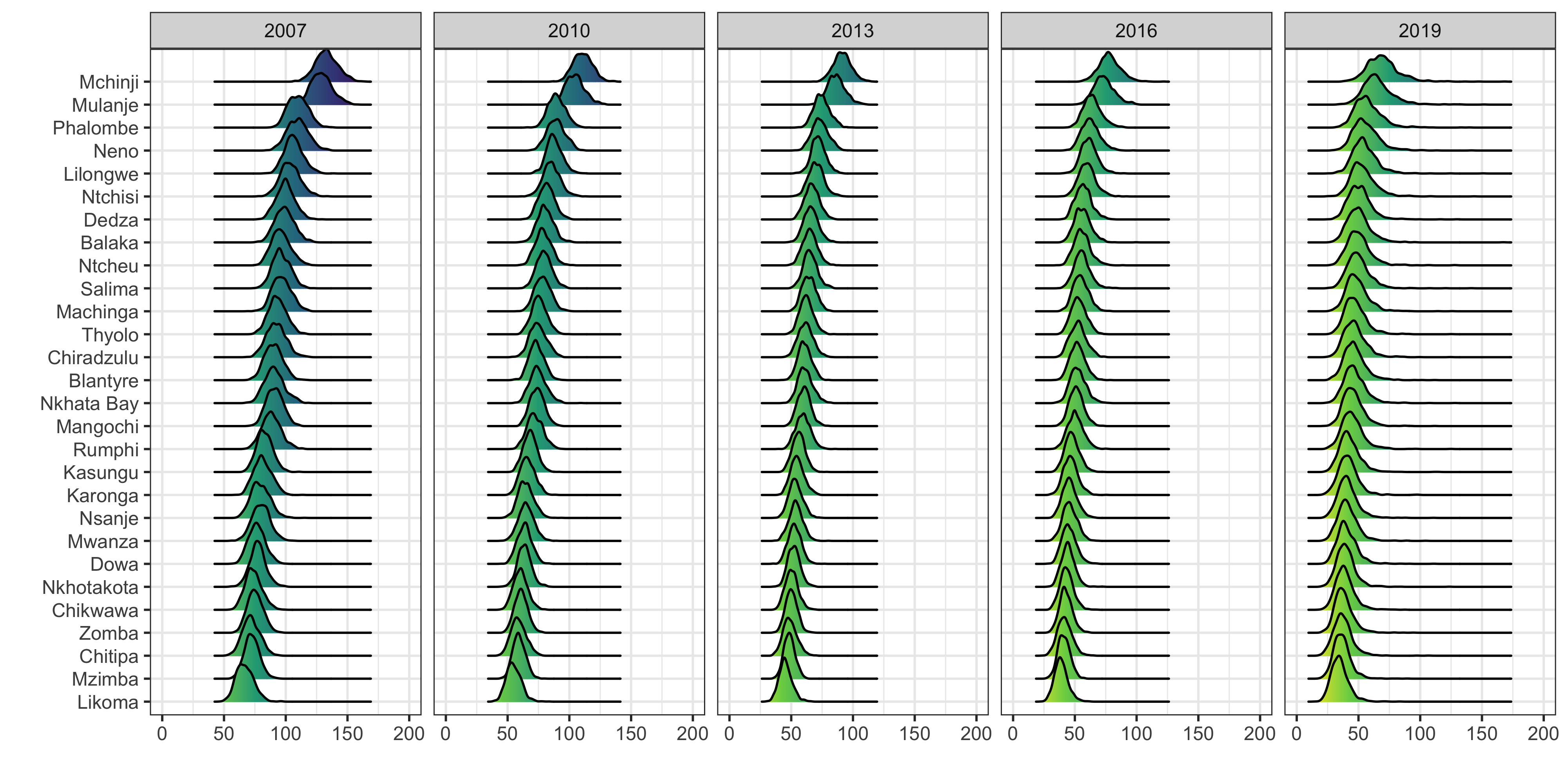} \caption{Posterior densities of the subnational estimates of U5MR using the 2015--2016 DHS in Malawi over selected years. Admin-2 regions are ordered by their median estimates in 2019. Estimates for 2019 in the last column are from the model projection and thus have higher uncertainty.}\label{fig:vis-4}
\end{figure}

The \texttt{ridgePlot()} function provides another visual comparison of the estimates and their associated uncertainty. Figure \ref{fig:vis-4} shows one such example where the marginal posterior densities of the estimates in the selected years are plotted with regions sorted by their posterior medians in the last plotted period. The posterior densities can also be grouped with all estimates in each region plotted in the same panel using \texttt{by.year\ =\ FALSE} argument in the \texttt{ridgePlot()} function. These plots are particularly useful to quickly identify regions with high and low estimates, while also showing the uncertainties associated with the rankings as well. The ranking of areas is often an important endeavor, since it can inform interventions in areas that are performing poorly or, more optimistically, allow areas with better outcomes to be examined to see if covariates (for example) are explaining their more positive performance.

\begin{verbatim}
ridgePlot(draws = est.bb, Amat = MalawiGraph, year.plot = year.plot,
          ncol = 5, per1000 = TRUE, order = -1, direction = -1) + xlim(c(0, 200))
\end{verbatim}

Finally, as models get more complicated, it becomes increasingly important to examine the estimated random effects for idiosyncratic behavior that may be evidence of model misspecification. The \CRANpkg{SUMMER} package provides tools to easily extract and plot posterior marginal distributions for each of the random effect components. We can use the \texttt{getDiag()} function to extract the posterior marginal distributions of the spatial, temporal, and space-time interaction terms from the fitted models, which can then be plotted in the similar fashion as the estimates. The space-time interaction term in the fitted model contains a sum of a region-specific linear trend and an AR(1) random effect. Thus we need posterior samples to compute their marginal distributions. We can feed the saved posterior draws from the \texttt{est.bb} here to speed up the computation.

\begin{verbatim}
r.time <- getDiag(fit.bb, field = "time")
r.space <- getDiag(fit.bb, field = "space")
r.interact <- getDiag(fit.bb, field = "spacetime", draws = est.bb$draws)
\end{verbatim}

\begin{figure}[!ht]
\includegraphics[width=1\linewidth,]{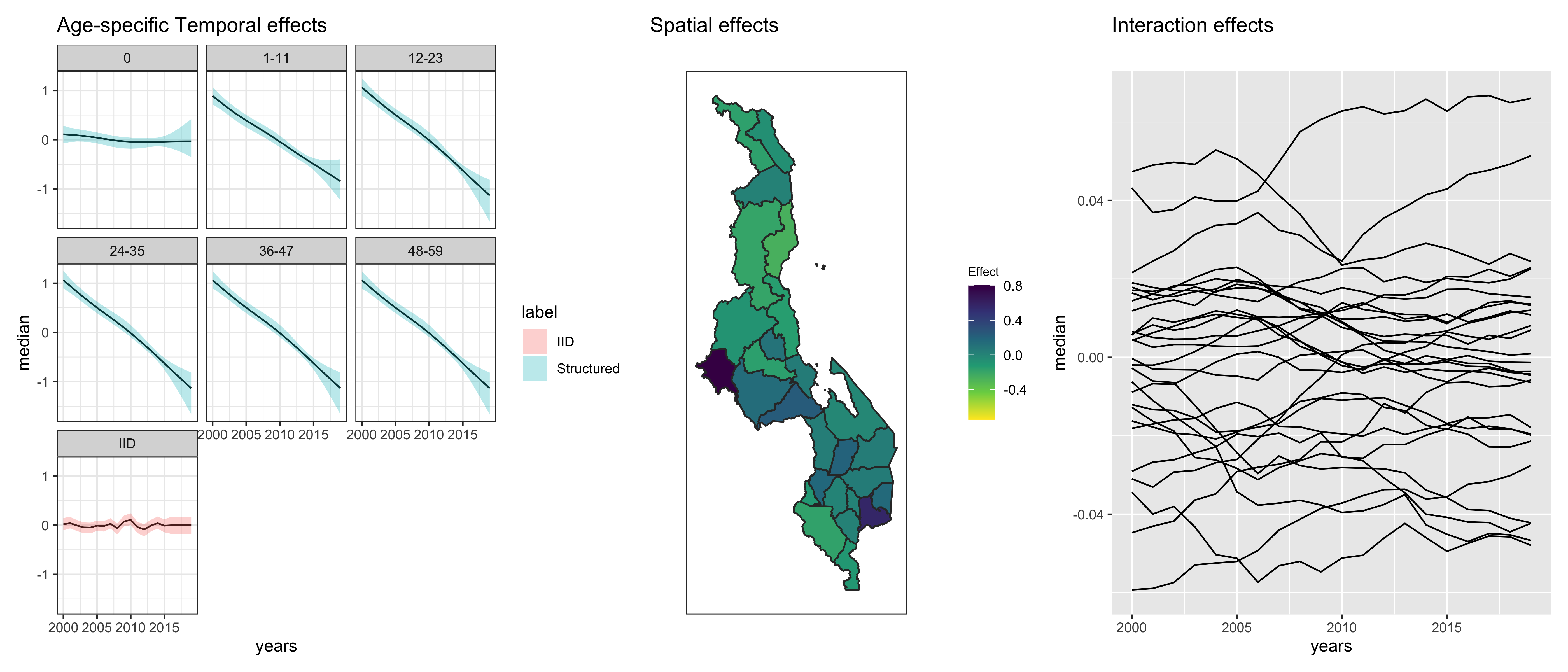} \caption{Posterior medians for the random effect terms in the cluster-level model using Malawi 2015--2016 DHS. Left: posterior medians and 95\% credible intervals of the age-specific temporal effects and the IID temporal shocks. Middle: posterior medians of the spatial effects. Right: posterior medians of the space-time interaction effects.}\label{fig:check-2}
\end{figure}

The extracted posterior summaries of the random effects can then be examined and visualized. Figure \ref{fig:check-2} shows the posterior summaries of the temporal, spatial, and interaction terms in the model.

\begin{verbatim}
g.time <- ggplot(r.time, aes(x = years, y = median, ymin=lower, ymax=upper)) + 
                 geom_line() + 
                 geom_ribbon(color=NA, aes(fill = label), alpha = 0.3) +
                 facet_wrap(~group, ncol = 3) + 
                 theme_bw() + 
                 ggtitle("Age-specific Temporal effects")
g.space <- mapPlot(subset(r.space, label = "Total"), 
                   geo=MalawiMap, by.data="region", by.geo = "ADM2_EN", 
                   direction = -1, variables="median", 
                   removetab=TRUE, legend.label = "Effect") + 
           ggtitle("Spatial effects") 
g.interact <- ggplot(r.interact, aes(x = years, y = median, group=region)) + 
                     geom_line() +  ggtitle("Interaction effects") 
g.time + g.space + g.interact + plot_layout(widths = c(3, 2, 3)) 
\end{verbatim}

\hypertarget{discussion}{%
\section{Discussion}\label{discussion}}

The present paper aims to provide a general overview of the R package \CRANpkg{SUMMER} for space-time smoothing of demographic and health indicators. The particular focus of this paper is on mortality estimation and the demonstration of the workflow for practitioners to fit flexible Bayesian smoothing models with DHS data. The implementation using \pkg{INLA} allows fast computation of these smoothing models. The Fay-Herriot estimates can usually be fit within seconds to minutes depending on the number of regions and time period. The cluster-level model may require longer computation time, especially with surveys containing many samples. We leave the fitting of more time-consuming models in the supplementary materials.

The \CRANpkg{SUMMER} package is in constant development. This paper introduced the core functionalities of the package. There are many more functionalities to tackle application-specific issues, such as different age-time interactions, aggregation over urban/rural strata, benchmarking to external national estimates, etc. We are also expanding the package to include more options for the traditional SAE models with fast implementations built on our current computational framework. Recent extensions built on \CRANpkg{SUMMER} functionalities include the recent addition of SAE methods in the \CRANpkg{survey} package (Lumley 2024) and the \CRANpkg{surveyPrev} package (Q. Dong et al. 2024) which provides a general pipeline to download, process, and map a broad variety of DHS indicators.

In the future, we have several plans to improve the functionality of \CRANpkg{SUMMER}. In the cluster-level model, we would like to allow different overdispersion parameters for different age groups. We plan to incorporate methods for child mortality estimation using summary birth history data (Hill et al. 2015; Wilson and Wakefield 2021) in which women provide only information on the number of children born, and number died, without the dates of these events. We also expect to extend the core functionalities to model other demographic and health indicators such as fertility. In our examples in this paper, we did not include covariates. In both the area-level and the cluster-level models, covariates can be included, see Wakefield, Okonek, and Pedersen (2020) for an example in the context of HIV prevalence mapping in Malawi. Finally, in the long term, we would like to incorporate continuous spatial models as well.

\hypertarget{references}{%
\section*{References}\label{references}}
\addcontentsline{toc}{section}{References}

\hypertarget{refs}{}
\begin{CSLReferences}{1}{0}
\leavevmode\vadjust pre{\hypertarget{ref-alkema_new_14}{}}%
Alkema, Leontine, and Jin Rou New. 2014. {``Global Estimation of Child Mortality Using a {B}ayesian {B}-Spline Bias-Reduction Model.''} \emph{The Annals of Applied Statistics} 8: 2122--49.

\leavevmode\vadjust pre{\hypertarget{ref-besagbayesian}{}}%
Besag, Julian, Jeremy York, and Annie Mollié. 1991. {``Bayesian Image Restoration, with Two Applications in Spatial Statistics (with Discussion).''} \emph{Annals of the Institute of Statistical Mathematics} 43: 1--59.

\leavevmode\vadjust pre{\hypertarget{ref-binder_83}{}}%
Binder, David A. 1983. {``On the Variances of Asymptotically Normal Estimators from Complex Surveys.''} \emph{International Statistical Review} 51: 279--92.

\leavevmode\vadjust pre{\hypertarget{ref-hbsae}{}}%
Boonstra, Harm Jan. 2012. \emph{Hbsae: Hierarchical Bayesian Small Area Estimation}.

\leavevmode\vadjust pre{\hypertarget{ref-surveyPrev}{}}%
Dong, Qianyu, Zehang R Li, Yunhan Wu, Andrea Boskovic, and Jon Wakefield. 2024. \emph{{surveyPrev}: Mapping the Prevalence of Binary Indicators Using Survey Data in Small Areas}. \url{https://CRAN.R-project.org/package=surveyPrev}.

\leavevmode\vadjust pre{\hypertarget{ref-dong_wakefield_20}{}}%
Dong, Tracy Qi, and Jon Wakefield. 2021. {``Modeling and Presentation of Vaccination Coverage Estimates Using Data from Household Surveys.''} \emph{Vaccine} 39 (18): 2584--94.

\leavevmode\vadjust pre{\hypertarget{ref-fay_herriot_79}{}}%
Fay, Robert E., and Roger A. Herriot. 1979. {``Estimates of Income for Small Places: An Application of {James--Stein} Procedure to Census Data.''} \emph{Journal of the American Statistical Association} 74: 269--77.

\leavevmode\vadjust pre{\hypertarget{ref-fuglstad2021two}{}}%
Fuglstad, Geir-Arne, Zehang Richard Li, and Jon Wakefield. 2021. {``The Two Cultures for Prevalence Mapping: Small Area Estimation and Spatial Statistics.''} \emph{arXiv Preprint arXiv:2110.09576}.

\leavevmode\vadjust pre{\hypertarget{ref-areas2012gadm}{}}%
Global Administrative Areas. 2012. {``GADM Database of Global Administrative Areas.''}

\leavevmode\vadjust pre{\hypertarget{ref-hajek_71}{}}%
Hájek, Jaroslav. 1971. {``Discussion of, {`{A}n Essay on the Logical Foundations of Survey Sampling, Part {I},'} by {D}. {B}asu.''} In \emph{Foundations of Statistical Inference}, edited by V. P. Godambe and D. A. Sprott. Toronto: Holt, Rinehart; Winston.

\leavevmode\vadjust pre{\hypertarget{ref-hill_etal_15}{}}%
Hill, Kenneth, Eoghan Brady, Linnea Zimmerman, Livia Montana, Romesh Silva, and Agbessi Amouzou. 2015. {``Monitoring Change in Child Mortality Through Household Surveys.''} \emph{PloS One} 10: e0137713.

\leavevmode\vadjust pre{\hypertarget{ref-horvitz_thompson_52}{}}%
Horvitz, Daniel G, and Donovan J Thompson. 1952. {``A Generalization of Sampling Without Replacement from a Finite Universe.''} \emph{Journal of the American Statistical Association} 47: 663--85.

\leavevmode\vadjust pre{\hypertarget{ref-knorrheld_00}{}}%
Knorr-Held, Leonhard. 2000. {``Bayesian Modelling of Inseparable Space-Time Variation in Disease Risk.''} \emph{Statistics in Medicine} 19: 2555--67.

\leavevmode\vadjust pre{\hypertarget{ref-li_etal_19}{}}%
Li, Zehang R, Yuan Hsiao, Jessica Godwin, Bryan D Martin, Jon Wakefield, and Samuel J Clark. 2019. {``Changes in the Spatial Distribution of the Under Five Mortality Rate: Small-Area Analysis of 122 {DHS} Surveys in 262 Subregions of 35 Countries in {A}frica.''} \emph{{PLoS One}}.

\leavevmode\vadjust pre{\hypertarget{ref-lumley2004analysis}{}}%
Lumley, Thomas. 2004. {``Analysis of Complex Survey Samples.''} \emph{Journal of Statistical Software} 9 (1): 1--19.

\leavevmode\vadjust pre{\hypertarget{ref-survey44}{}}%
---------. 2024. {``Survey: Analysis of Complex Survey Samples.''}

\leavevmode\vadjust pre{\hypertarget{ref-macfeely20}{}}%
MacFeely, Steve. 2020. {``Measuring the Sustainable Development Goal Indicators: An Unprecedented Statistical Challenge.''} \emph{Journal of Official Statistics} 36 (2): 361--78.

\leavevmode\vadjust pre{\hypertarget{ref-mercer_etal_15}{}}%
Mercer, Laina D, Jon Wakefield, Athena Pantazis, Angelina M Lutambi, Honorati Masanja, and Samuel Clark. 2015. {``Small Area Estimation of Childhood Mortality in the Absence of Vital Registration.''} \emph{Annals of Applied Statistics} 9: 1889--1905.

\leavevmode\vadjust pre{\hypertarget{ref-molina_marhuenda_15}{}}%
Molina, Isabel, and Yolanda Marhuenda. 2015. {``\pkg{sae}: An \proglang{R} Package for Small Area Estimation.''} \emph{The R Journal} 7: 81--98.

\leavevmode\vadjust pre{\hypertarget{ref-paige_etal_20}{}}%
Paige, John, Geir-Arne Fuglstad, Andrea Riebler, and Jon Wakefield. 2020. {``Model-Based Approaches to Analysing Spatial Data from Complex Surveys.''} \emph{Journal of Survey Statistics and Methodology}.

\leavevmode\vadjust pre{\hypertarget{ref-pedersenandliu_2012}{}}%
Pedersen, Jon, and Jing Liu. 2012. {``Child Mortality Estimation: Appropriate Time Periods for Child Mortality Estimates from Full Birth Histories.''} \emph{PLoS Med} 9 (8): e1001289.

\leavevmode\vadjust pre{\hypertarget{ref-patchwork}{}}%
Pedersen, Thomas Lin. 2019. \emph{\pkg{patchwork}: The Composer of Plots}. \url{https://CRAN.R-project.org/package=patchwork}.

\leavevmode\vadjust pre{\hypertarget{ref-permatasari2021msae}{}}%
Permatasari, Novia, and Azka Ubaidillah. 2021. {``{msae}: An {R} Package of Multivariate {Fay Herriot} Models for Small Area Estimation.''} \emph{The R Journal}.

\leavevmode\vadjust pre{\hypertarget{ref-rao_molina_15}{}}%
Rao, John NK, and Isabel Molina. 2015. \emph{Small Area Estimation, Second Edition}. New York: John Wiley.

\leavevmode\vadjust pre{\hypertarget{ref-riebler_etal_16}{}}%
Riebler, Andrea, Sigrunn H Sørbye, Daniel Simpson, and Håvard Rue. 2016. {``An Intuitive {B}ayesian Spatial Model for Disease Mapping That Accounts for Scaling.''} \emph{Statistical Methods in Medical Research} 25: 1145--65.

\leavevmode\vadjust pre{\hypertarget{ref-rue_knorrheld_05}{}}%
Rue, Håvard, and Leonhard Held. 2005. \emph{{G}aussian {M}arkov {R}andom {F}ields: {T}heory and {A}pplication}. Boca Raton: Chapman; Hall/CRC Press.

\leavevmode\vadjust pre{\hypertarget{ref-rue_etal_09}{}}%
Rue, H., S. Martino, and N. Chopin. 2009. {``Approximate {B}ayesian Inference for Latent {G}aussian Models Using Integrated Nested {L}aplace Approximations (with Discussion).''} \emph{Journal of the Royal Statistical Society, Series B} 71: 319--92.

\leavevmode\vadjust pre{\hypertarget{ref-schluter2021space}{}}%
Schlüter, Benjamin-Samuel, and Bruno Masquelier. 2021. {``Space-Time Smoothing of Mortality Estimates in Children Aged 5-14 in Sub-Saharan Africa.''} \emph{Plos One} 16 (1): e0245596.

\leavevmode\vadjust pre{\hypertarget{ref-rsae}{}}%
Schoch, Tobias. 2014. \emph{Rsae: Robust Small Area Estimation}.

\leavevmode\vadjust pre{\hypertarget{ref-BayesSAE}{}}%
Shi, Chengchun. 2018. \emph{{BayesSAE}: Bayesian Analysis of Small Area Estimation}.

\leavevmode\vadjust pre{\hypertarget{ref-dhsspatial}{}}%
The DHS Program. 2020. {``Spatial Data Repository.''} \url{https://spatialdata.dhsprogram.com/home/}.

\leavevmode\vadjust pre{\hypertarget{ref-un2023}{}}%
United Nations Inter-agency Group for Child Mortality Estimation. 2023. {``Subnational Under-Five and Neonatal Mortality Estimates, 2000--2021 Estimates Developed by the {United Nations Inter-agency Group for Child Mortality Estimation}.''} \emph{United Nations Children's Fund, New York}.

\leavevmode\vadjust pre{\hypertarget{ref-wakefield_etal_19}{}}%
Wakefield, Jon, Geir-Arne Fuglstad, Andrea Riebler, Jessica Godwin, Katie Wilson, and Samuel Clark. 2019. {``Estimating Under Five Mortality in Space and Time in a Developing World Context.''} \emph{Statistical Methods in Medical Research} 28: 2614--34.

\leavevmode\vadjust pre{\hypertarget{ref-wakefield_okonek_pedersen_20}{}}%
Wakefield, Jon, Taylor Okonek, and Jon Pedersen. 2020. {``Small Area Estimation of Health Outcomes.''} \emph{International Statistical Review}.

\leavevmode\vadjust pre{\hypertarget{ref-walker_etal_12}{}}%
Walker, Neff, Kenneth Hill, and Fengmin Zhao. 2012. {``Child Mortality Estimation: Methods Used to Adjust for Bias Due to {AIDS} in Estimating Trends in Under-Five Mortality.''} \emph{PLoS Med} 9: e1001298.

\leavevmode\vadjust pre{\hypertarget{ref-rdhs}{}}%
Watson, OJ, and Jeff Eaton. 2019. \emph{\pkg{rdhs}: {API} Client and Dataset Management for the {Demographic and Health Survey (DHS)} Data}. \url{https://CRAN.R-project.org/package=rdhs}.

\leavevmode\vadjust pre{\hypertarget{ref-wickham_ggplot2}{}}%
Wickham, Hadley. 2016. \emph{\pkg{ggplot2}: Elegant Graphics for Data Analysis}. Springer-Verlag New York. \url{https://ggplot2.tidyverse.org}.

\leavevmode\vadjust pre{\hypertarget{ref-wilson2021child}{}}%
Wilson, Katie, and Jon Wakefield. 2021. {``Child Mortality Estimation Incorporating Summary Birth History Data.''} \emph{Biometrics} 77 (4): 1456--66.

\leavevmode\vadjust pre{\hypertarget{ref-wolter_07}{}}%
Wolter, Kirk. 2007. \emph{Introduction to Variance Estimation}. Springer Science \& Business Media.

\leavevmode\vadjust pre{\hypertarget{ref-wu2021spatial}{}}%
Wu, Yunhan, Zehang Richard Li, Benjamin K. Mayala, Houjie Wang, Peter Gao, Johnny Paige, Geir-Arne Fuglstad, et al. 2021. {``Spatial Modeling for Subnational Administrative Level 2 Small-Area Estimation.''} \emph{DHS Spatial Analysis Reports No. 21. Rockville, Maryland, USA: ICF.}

\end{CSLReferences}

\address{%
Zehang Richard Li\\
University of California, Santa Cruz\\%
Santa Cruz CA, USA\\
\textit{ORCiD: \href{https://orcid.org/0000-0002-9079-593X}{0000-0002-9079-593X}}\\%
\href{mailto:lizehang@ucsc.edu}{\nolinkurl{lizehang@ucsc.edu}}%
}

\address{%
Bryan D Martin\\
University of Washington\\%
Seattle, WA, USA\\
\href{mailto:bmartin6@uw.edu}{\nolinkurl{bmartin6@uw.edu}}%
}

\address{%
Tracy Qi Dong\\
Fred Hutchinson Cancer Research Center\\%
Seattle, WA, USA\\
\href{mailto:qd8@uw.edu}{\nolinkurl{qd8@uw.edu}}%
}

\address{%
Geir-Arne Fuglstad\\
Norwegian University of Science and Technology\\%
Trondheim, Norway\\
\href{mailto:geir-arne.fuglstad@ntnu.no}{\nolinkurl{geir-arne.fuglstad@ntnu.no}}%
}

\address{%
Jessica Godwin\\
University of Washington\\%
Seattle, WA, USA\\
\href{mailto:jlg0003@uw.edu}{\nolinkurl{jlg0003@uw.edu}}%
}

\address{%
John Paige\\
Norwegian University of Science and Technology\\%
Trondheim, Norway\\
\href{mailto:paigejo@gmail.com}{\nolinkurl{paigejo@gmail.com}}%
}

\address{%
Andrea Riebler\\
Norwegian University of Science and Technology\\%
Trondheim, Norway\\
\href{mailto:andrea.riebler@ntnu.no}{\nolinkurl{andrea.riebler@ntnu.no}}%
}

\address{%
Samuel J Clark\\
The Ohio State University\\%
Columbus, OH, USA\\
\href{mailto:work@samclark.net}{\nolinkurl{work@samclark.net}}%
}

\address{%
Jon Wakefield\\
University of Washington\\%
Seattle, WA, USA\\
\href{mailto:jonno@uw.edu}{\nolinkurl{jonno@uw.edu}}%
}

\end{article}

\newpage
\section*{Appendix}
\appendix

\renewcommand\thesection{\Alph{section}}

\section{Prior specification}\label{sec:model-prior}

In all the model implementations, we apply penalised complexity (PC) priors to model the random effects \citep{simpson_etal_17}. These priors are proper and parameterization invariant. The basis of PC priors is to regard each model component as a flexible extension of a so-called base model. Considering an unstructured iid model component, the base model would be to remove this component from the linear predictor by letting its variance parameter go to zero. This is also the base model for any simple Gaussian model component with mean zero. The main idea is to follow Occam's razor and favor less complex, or more intuitive, models unless the data suggest otherwise. Of note, state-of-the-art priors, such as the inverse gamma prior for a variance parameter, put zero density mass at the base model and as such do not allow the recovery of this model. The PC prior is specified by following a number of desirable principles and is derived based on the Kullback-Leibler distance of the flexible model from the base model. For details we refer to \citet{simpson_etal_17}. Here, we will shortly comment on the PC priors relevant for the parameters used in our models and their default hyperparameters. All the PC priors can be specified by the user in the function calls using arguments such as \texttt{pc.u} and \texttt{pc.alpha}.

Considering a simple Gaussian model component with standard deviation parameter \(\sigma\), the PC prior results in an exponential distribution for \(\sigma\). The rate parameter \(\lambda_\sigma\) can be informed using a probability contrast of the form \(\mbox{Prob}(\sigma > U_\sigma) = \alpha_\sigma\), which leads to \(\lambda_\sigma = -\log(\alpha_\sigma)/ U_\sigma\) \citep{simpson_etal_17}. The \pkg{SUMMER} package uses as default \(U_\sigma=1\) and \(\alpha_\sigma=0.01\), which means that the 99th percentile of the prior is at 1.

For the structured spatial random effects, we use the BYM2 model \citep{riebler_etal_16, simpson_etal_17}. It has a structured and unstructured term, and uses a single variance, \(\sigma^2\), that represents the marginal spatial variance and a mixing parameter \(\phi \in [0,1]\) specifying the proportion of spatial variation. To interpret \(\sigma\) as a marginal standard deviation, the spatial component needs to be scaled, so that \(\mbox{Var}(e_i) \approx \mbox{Var}(S_i) \approx 1\). This leads to:
\begin{equation*}
  \mathbf{e} + \bS = \sigma (\sqrt{(1-\phi)}\mathbf{e} ^\star+ \sqrt{\phi} \bS ^\star)
\end{equation*}
where \(\mathbf{e}^\star\) is iid normally distributed with fixed variance equal to 1 and \(\bS^\star\) is the scaled ICAR model. We follow \cite{riebler_etal_16} and scale the ICAR component so that the geometric mean of the marginal variances of \(S_i\) is equal to 1. Note that we also apply this scaling procedure to all intrinsic model components, such as random walk of order 1 or 2 components \citep{sorbye-rue-2014}, to ensure interpretability of the prior distributions assigned to their flexibility parameters. The BYM2 model has a two-stage base model, with the first implying the absence of any spatial effect by setting \(\sigma\) equal to zero, and the second by assuming \(\phi=0\) and therefore only unstructured spatial variation. For \(\sigma\) we use an exponential prior as outlined before. The prior for \(\phi\) depends on the study-specific neighborhood graph and is not available in closed form, see \cite{riebler_etal_16} for details. Its hyperparameter \(\lambda_{\phi}\) can be derived from
\(\mbox{Prob}(\phi < U_\phi) =\alpha_\phi\). The \pkg{SUMMER} package uses as default \(U_\phi=0.5\) and \(\alpha_\phi=2/3\), which means that the 66.6th percentile of the prior is at 0.5, so that values of \(\phi\) less than \(0.5\) are preferred a little more, a priori.

For the autoregressive model for time effects, we again use an exponential prior for the marginal standard deviation. For the autocorrelation correlation coefficient \(\omega\), we assume as base model \(\omega = 1\). This represents a limiting random walk which assumes that the process does not change in time. The prior for \(\omega\) is again not available in closed form, see \cite{sorbye_rue_17} for details. Its hyperparameter \(\lambda_{\omega}\) can be found from
\(\mbox{Prob}(\omega > U_\omega) =\alpha_\omega\). The \pkg{SUMMER} package uses as default \(U_\omega=0.7\) and \(\alpha_\omega=0.9\), which means that the 10th percentile of the prior is at 0.7, and therefore preferring values of \(\phi\) that are close to 1.

The space-time interaction terms, \(\delta_{it}\), are modeled with the Type I, II, III, IV models of \citet{knorrheld_00}. The Type I model assumes iid interaction terms, the Type II model that the interactions are temporally structured but independent in space and the Type III model that the interactions are iid in time but spatially structured via an ICAR model. For the default Type IV interaction, we assume the specified temporal model and spatial (ICAR) structured effects interact. When the temporal component in the space-time interaction terms are modeled with a random walk of order 1 or an autoregressive model of order 1, we may also allow area-specific deviations from the main temporal trends by letting \(\delta_{it} = b_i t + \delta_{it}^\star\), where \(\delta_{it}^\star\) follows the specified interaction model and \(b_i\) are random slopes. This allows us to capture more flexible temporal dynamics, and may aid in area-specific predictions. The random slopes are modeled with a Gaussian prior. To facilitate interpretation, we scale the time index to be from \(-0.5\) to \(0.5\), so that the random slope can be interpreted as the total deviation from the main time trend from the first and last years to be projected, on the logit scale. Users can specify priors for the random slopes with the PC prior so that \(\mbox{Prob}(|b| < U_b) =\alpha_b\) using arguments \texttt{pc.st.slope.u} and \texttt{pc.st.slope.alpha}.

\section{Additional analysis of the simulated dataset}

\subsection{Unstratified cluster-level model}

In this section, we present additional analysis for the simulated dataset in Example 2 of the main paper. We first load the package, data, and compute the direct estimates of U5MR following the example in the main paper.

\begin{Shaded}
\begin{Highlighting}[]
\FunctionTok{library}\NormalTok{(SUMMER)}
\FunctionTok{library}\NormalTok{(stringr)}
\FunctionTok{library}\NormalTok{(dplyr)}
\FunctionTok{library}\NormalTok{(ggplot2)}
\FunctionTok{library}\NormalTok{(patchwork)}
\FunctionTok{data}\NormalTok{(DemoData)}

\NormalTok{periods }\OtherTok{\textless{}{-}} \FunctionTok{c}\NormalTok{(}\StringTok{"85{-}89"}\NormalTok{, }\StringTok{"90{-}94"}\NormalTok{, }\StringTok{"95{-}99"}\NormalTok{, }\StringTok{"00{-}04"}\NormalTok{, }\StringTok{"05{-}09"}\NormalTok{, }\StringTok{"10{-}14"}\NormalTok{)}
\NormalTok{directU5 }\OtherTok{\textless{}{-}} \FunctionTok{getDirectList}\NormalTok{(}\AttributeTok{births =}\NormalTok{ DemoData, }\AttributeTok{years =}\NormalTok{ periods, }
                          \AttributeTok{regionVar =} \StringTok{"region"}\NormalTok{, }\AttributeTok{timeVar =} \StringTok{"time"}\NormalTok{, }
                          \AttributeTok{clusterVar =} \StringTok{"\textasciitilde{}clustid + id"}\NormalTok{, }\AttributeTok{ageVar =} \StringTok{"age"}\NormalTok{, }
                          \AttributeTok{weightsVar =} \StringTok{"weights"}\NormalTok{)}
\end{Highlighting}
\end{Shaded}

We now describe the fitting of the cluster level model. For simplicity, we first assume the survey was designed so that each of the four regions was a strata (thus no additional stratification within regions). The stratified analysis is left to the next section.

First, we calculate the number of person-months and number of deaths for each cluster, time period, and age group. Notice that we do not need to impute all the 0's for combinations that do not exist in the data. We first create the data frame using the \texttt{getCounts()} function. The \texttt{getCounts()} function prepares the aggregated count dataset without modifying the original column names. For the model fitting functions to correctly identify the data columns, we rename the cluster ID and time period columns to be `cluster' and `years'. The response variable is `Y' and the binomial total is `total'.

\begin{Shaded}
\begin{Highlighting}[]
\NormalTok{counts.all }\OtherTok{\textless{}{-}} \ConstantTok{NULL}
\ControlFlowTok{for}\NormalTok{(i }\ControlFlowTok{in} \DecValTok{1}\SpecialCharTok{:}\FunctionTok{length}\NormalTok{(DemoData))\{}
\NormalTok{    vars }\OtherTok{\textless{}{-}} \FunctionTok{c}\NormalTok{(}\StringTok{"clustid"}\NormalTok{, }\StringTok{"region"}\NormalTok{, }\StringTok{"time"}\NormalTok{, }\StringTok{"age"}\NormalTok{)}
\NormalTok{    counts }\OtherTok{\textless{}{-}} \FunctionTok{getCounts}\NormalTok{(DemoData[[i]][, }\FunctionTok{c}\NormalTok{(vars, }\StringTok{"died"}\NormalTok{)], }\AttributeTok{variables =} \StringTok{\textquotesingle{}died\textquotesingle{}}\NormalTok{, }
                        \AttributeTok{by =}\NormalTok{ vars, }\AttributeTok{drop=}\ConstantTok{TRUE}\NormalTok{)}
\NormalTok{    counts }\OtherTok{\textless{}{-}}\NormalTok{ counts }\SpecialCharTok{\%\textgreater{}\%} \FunctionTok{mutate}\NormalTok{(}\AttributeTok{cluster =}\NormalTok{ clustid, }\AttributeTok{years =}\NormalTok{ time, }\AttributeTok{Y=}\NormalTok{died)}
\NormalTok{    counts}\SpecialCharTok{$}\NormalTok{survey }\OtherTok{\textless{}{-}} \FunctionTok{names}\NormalTok{(DemoData)[i] }
\NormalTok{    counts.all }\OtherTok{\textless{}{-}} \FunctionTok{rbind}\NormalTok{(counts.all, counts)}
\NormalTok{\}}
\FunctionTok{head}\NormalTok{(counts.all)}
\end{Highlighting}
\end{Shaded}

\begin{verbatim}
##   clustid  region  time age died total cluster years Y survey
## 1      36 central 85-89   0    0     1      36 85-89 0   1999
## 2      38 central 85-89   0    0     1      38 85-89 0   1999
## 3      91 central 85-89   0    0     1      91 85-89 0   1999
## 4     101 central 85-89   0    0     1     101 85-89 0   1999
## 5     128 central 85-89   0    0     1     128 85-89 0   1999
## 6     129 central 85-89   0    0     1     129 85-89 0   1999
\end{verbatim}

With the created data frame, we fit the cluster-level model using the \texttt{smoothCluster()} function and obtain the estimates with the \texttt{getSmoothed()} function. Notice that here we need to specify the age groups (\texttt{age.groups}), the length of each age group (\texttt{age.n}) in months, and how the age groups are mapped to the temporal random effects (\texttt{age.rw.group}). In the default case, \texttt{age.rw.group\ =\ c(1,\ 2,\ 3,\ 3,\ 3,\ 3)} means the first two age groups each has its own temporal trend, the the following four age groups share the same temporal trend. We start with the default temporal model of random walk or order 2 on the 5-year periods in this dataset (with real data, we can use a finer temporal resolution). We add survey iid effects to the model as well using \texttt{survey.effect\ =\ TRUE} argument.

\begin{Shaded}
\begin{Highlighting}[]
\NormalTok{fit.bb.sim }\OtherTok{\textless{}{-}} \FunctionTok{smoothCluster}\NormalTok{(}\AttributeTok{data =}\NormalTok{ counts.all, }\AttributeTok{Amat =}\NormalTok{ DemoMap}\SpecialCharTok{$}\NormalTok{Amat, }
                \AttributeTok{family =} \StringTok{"betabinomial"}\NormalTok{,}
                \AttributeTok{year.label =} \FunctionTok{c}\NormalTok{(periods, }\StringTok{"15{-}19"}\NormalTok{), }
                \AttributeTok{age.group =} \FunctionTok{c}\NormalTok{(}\StringTok{"0"}\NormalTok{, }\StringTok{"1{-}11"}\NormalTok{, }\StringTok{"12{-}23"}\NormalTok{, }\StringTok{"24{-}35"}\NormalTok{, }\StringTok{"36{-}47"}\NormalTok{, }\StringTok{"48{-}59"}\NormalTok{),}
                \AttributeTok{age.n =} \FunctionTok{c}\NormalTok{(}\DecValTok{1}\NormalTok{, }\DecValTok{11}\NormalTok{, }\DecValTok{12}\NormalTok{, }\DecValTok{12}\NormalTok{, }\DecValTok{12}\NormalTok{, }\DecValTok{12}\NormalTok{),}
                \AttributeTok{age.time.group =} \FunctionTok{c}\NormalTok{(}\DecValTok{1}\NormalTok{, }\DecValTok{2}\NormalTok{, }\DecValTok{3}\NormalTok{, }\DecValTok{3}\NormalTok{, }\DecValTok{3}\NormalTok{, }\DecValTok{3}\NormalTok{),}
                \AttributeTok{time.model =} \StringTok{"rw2"}\NormalTok{,}
                \AttributeTok{st.time.model =} \StringTok{"ar1"}\NormalTok{,}
                \AttributeTok{pc.st.slope.u =} \DecValTok{2}\NormalTok{, }\AttributeTok{pc.st.slope.alpha =} \FloatTok{0.1}\NormalTok{,  }
                \AttributeTok{survey.effect =} \ConstantTok{TRUE}\NormalTok{)}
\NormalTok{est.bb.sim }\OtherTok{\textless{}{-}} \FunctionTok{getSmoothed}\NormalTok{(fit.bb.sim, }\AttributeTok{nsim =} \DecValTok{1000}\NormalTok{)}
\end{Highlighting}
\end{Shaded}

\begin{figure}[!ht]
\includegraphics[width=1\linewidth,]{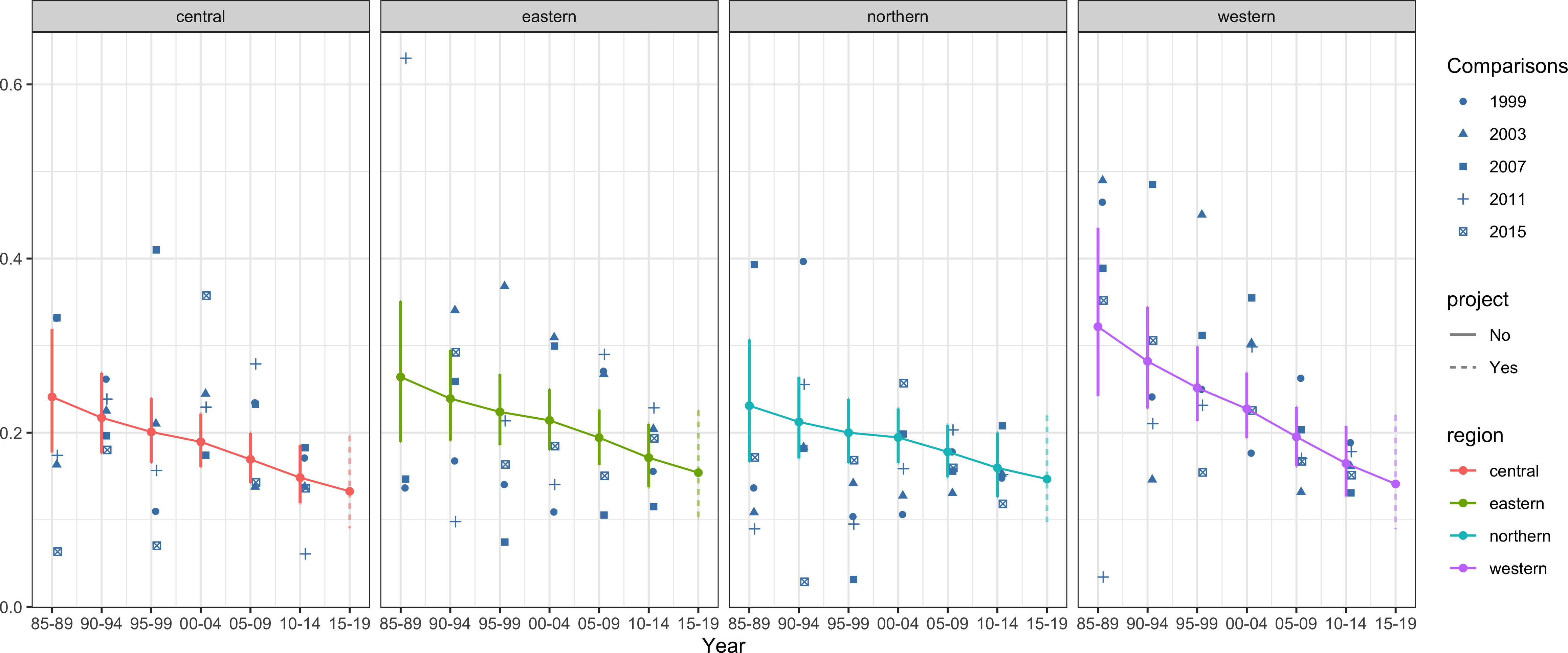} \caption{Cluster-level models estimates of subnational U5MR using multiple simulated surveys. The blue dots are direct estimates from each of the surveys.}\label{fig:sim-fit-3}
\end{figure}

We visualize the results from the cluster level model in Figure \ref{fig:sim-fit-3}. We also overlay the survey-specific direct estimates using the \texttt{data.add} argument.

\begin{Shaded}
\begin{Highlighting}[]
\FunctionTok{plot}\NormalTok{(est.bb.sim}\SpecialCharTok{$}\NormalTok{overall, }\AttributeTok{plot.CI=}\ConstantTok{TRUE}\NormalTok{,  }\AttributeTok{data.add =}\NormalTok{ directU5, }
            \AttributeTok{option.add =} \FunctionTok{list}\NormalTok{(}\AttributeTok{point =} \StringTok{"mean"}\NormalTok{, }\AttributeTok{by =} \StringTok{"surveyYears"}\NormalTok{), }
            \AttributeTok{color.add=}\StringTok{"steelblue"}\NormalTok{) }\SpecialCharTok{+} \FunctionTok{facet\_wrap}\NormalTok{(}\SpecialCharTok{\textasciitilde{}}\NormalTok{region, }\AttributeTok{ncol =} \DecValTok{4}\NormalTok{) }
\end{Highlighting}
\end{Shaded}

\subsection{Stratified cluster-level model}

We now describe the fitting of the cluster-level model for U5MR taking into account the urban/rural stratification. For this simulated dataset, the strata variable is coded as region crossed by urban/rural status. For our analysis with urban/rural stratified model, we first construct a new strata variable that contains only the urban/rural status, i.e., the additional stratification within each region and computes the counts of person-months annd death similar to the previous case.

\begin{Shaded}
\begin{Highlighting}[]
\ControlFlowTok{for}\NormalTok{(i }\ControlFlowTok{in} \DecValTok{1}\SpecialCharTok{:}\FunctionTok{length}\NormalTok{(DemoData))\{}
\NormalTok{  strata }\OtherTok{\textless{}{-}}\NormalTok{ DemoData[[i]]}\SpecialCharTok{$}\NormalTok{strata}
\NormalTok{  DemoData[[i]]}\SpecialCharTok{$}\NormalTok{strata[}\FunctionTok{grep}\NormalTok{(}\StringTok{"urban"}\NormalTok{, strata)] }\OtherTok{\textless{}{-}} \StringTok{"urban"} 
\NormalTok{  DemoData[[i]]}\SpecialCharTok{$}\NormalTok{strata[}\FunctionTok{grep}\NormalTok{(}\StringTok{"rural"}\NormalTok{, strata)] }\OtherTok{\textless{}{-}} \StringTok{"rural"}
\NormalTok{\}}
\NormalTok{counts.all }\OtherTok{\textless{}{-}} \ConstantTok{NULL}
\ControlFlowTok{for}\NormalTok{(i }\ControlFlowTok{in} \DecValTok{1}\SpecialCharTok{:}\FunctionTok{length}\NormalTok{(DemoData))\{}
\NormalTok{  vars }\OtherTok{\textless{}{-}} \FunctionTok{c}\NormalTok{(}\StringTok{"clustid"}\NormalTok{, }\StringTok{"region"}\NormalTok{, }\StringTok{"strata"}\NormalTok{, }\StringTok{"time"}\NormalTok{, }\StringTok{"age"}\NormalTok{)}
\NormalTok{  counts }\OtherTok{\textless{}{-}} \FunctionTok{getCounts}\NormalTok{(DemoData[[i]][, }\FunctionTok{c}\NormalTok{(vars, }\StringTok{"died"}\NormalTok{)], }\AttributeTok{variables =} \StringTok{\textquotesingle{}died\textquotesingle{}}\NormalTok{, }
              \AttributeTok{by =}\NormalTok{ vars, }\AttributeTok{drop=}\ConstantTok{TRUE}\NormalTok{)}
\NormalTok{  counts }\OtherTok{\textless{}{-}}\NormalTok{ counts }\SpecialCharTok{\%\textgreater{}\%} \FunctionTok{mutate}\NormalTok{(}\AttributeTok{cluster =}\NormalTok{ clustid, }\AttributeTok{years =}\NormalTok{ time, }\AttributeTok{Y=}\NormalTok{died)}
\NormalTok{  counts}\SpecialCharTok{$}\NormalTok{survey }\OtherTok{\textless{}{-}} \FunctionTok{names}\NormalTok{(DemoData)[i] }
\NormalTok{  counts.all }\OtherTok{\textless{}{-}} \FunctionTok{rbind}\NormalTok{(counts.all, counts)}
\NormalTok{\}}
\end{Highlighting}
\end{Shaded}

With the created data frame, we fit the cluster-level model using the \texttt{smoothCluster} function. The temporal main effects are defined for each stratum separately (specified by \texttt{strata.time.effect\ =\ TRUE}, so in total six random walks are used to model the main temporal effect.

\begin{Shaded}
\begin{Highlighting}[]
\NormalTok{fit.bb  }\OtherTok{\textless{}{-}} \FunctionTok{smoothCluster}\NormalTok{(}\AttributeTok{data =}\NormalTok{ counts.all, }\AttributeTok{Amat =}\NormalTok{ DemoMap}\SpecialCharTok{$}\NormalTok{Amat, }
            \AttributeTok{family =} \StringTok{"betabinomial"}\NormalTok{,}
            \AttributeTok{year.label =} \FunctionTok{c}\NormalTok{(periods, }\StringTok{"15{-}19"}\NormalTok{), }
            \AttributeTok{age.group =} \FunctionTok{c}\NormalTok{(}\StringTok{"0"}\NormalTok{, }\StringTok{"1{-}11"}\NormalTok{, }\StringTok{"12{-}23"}\NormalTok{, }\StringTok{"24{-}35"}\NormalTok{, }\StringTok{"36{-}47"}\NormalTok{, }\StringTok{"48{-}59"}\NormalTok{),}
            \AttributeTok{age.n =} \FunctionTok{c}\NormalTok{(}\DecValTok{1}\NormalTok{, }\DecValTok{11}\NormalTok{, }\DecValTok{12}\NormalTok{, }\DecValTok{12}\NormalTok{, }\DecValTok{12}\NormalTok{, }\DecValTok{12}\NormalTok{),}
            \AttributeTok{age.time.group =} \FunctionTok{c}\NormalTok{(}\DecValTok{1}\NormalTok{, }\DecValTok{2}\NormalTok{, }\DecValTok{3}\NormalTok{, }\DecValTok{3}\NormalTok{, }\DecValTok{3}\NormalTok{, }\DecValTok{3}\NormalTok{),}
            \AttributeTok{time.model =} \StringTok{"rw2"}\NormalTok{,}
            \AttributeTok{st.time.model =} \StringTok{"ar1"}\NormalTok{,}
            \AttributeTok{pc.st.slope.u =} \DecValTok{2}\NormalTok{, }\AttributeTok{pc.st.slope.alpha =} \FloatTok{0.1}\NormalTok{,  }
            \AttributeTok{survey.effect =} \ConstantTok{TRUE}\NormalTok{, }
            \AttributeTok{strata.time.effect =} \ConstantTok{TRUE}\NormalTok{)}
\end{Highlighting}
\end{Shaded}

\begin{verbatim}
## ----------------------------------
## Cluster-level model
##   Main temporal model:        rw2
##   Number of time periods:     7
##   Spatial effect:             bym2
##   Number of regions:          4
##   Interaction temporal model: ar1
##   Interaction type:           4
##   Interaction random slopes:  yes
##   Number of age groups: 6
##   Stratification: yes
##   Number of age-specific fixed effect intercept per stratum: 6
##   Number of age-specific trends per stratum: 3
##   Strata-specific temporal trends: yes
##   Survey effect: yes
## ----------------------------------
\end{verbatim}

\begin{Shaded}
\begin{Highlighting}[]
\NormalTok{est.bb }\OtherTok{\textless{}{-}} \FunctionTok{getSmoothed}\NormalTok{(fit.bb, }\AttributeTok{nsim =} \DecValTok{1000}\NormalTok{, }\AttributeTok{CI =} \FloatTok{0.95}\NormalTok{, }\AttributeTok{save.draws=}\ConstantTok{TRUE}\NormalTok{)}
\end{Highlighting}
\end{Shaded}

\begin{verbatim}
## Starting posterior sampling...
## Cleaning up results...
## No strata weights has been supplied. Overall estimates are not calculated.
\end{verbatim}

\begin{Shaded}
\begin{Highlighting}[]
\FunctionTok{summary}\NormalTok{(fit.bb)}
\end{Highlighting}
\end{Shaded}

\begin{verbatim}
## ----------------------------------
## Cluster-level model
##   Main temporal model:        rw2
##   Number of time periods:     7
##   Spatial effect:             bym2
##   Number of regions:          4
##   Interaction temporal model: ar1
##   Interaction type:           4
##   Interaction random slopes:  yes
##   Number of age groups: 6
##   Stratification: yes
##   Number of age group fixed effect intercept per stratum: 6
##   Number of age-specific trends per stratum: 3
##   Strata-specific temporal trends: yes
##   Survey effect: yes
## ----------------------------------
## Fixed Effects
##                          mean   sd 0.025quant 0.5quant 0.97quant mode kld
## age.intercept0:rural     -3.0 0.14       -3.3     -3.0      -2.7 -3.0   0
## age.intercept1-11:rural  -4.7 0.11       -4.9     -4.7      -4.5 -4.7   0
## age.intercept12-23:rural -5.8 0.14       -6.1     -5.8      -5.5 -5.8   0
## age.intercept24-35:rural -6.6 0.20       -6.9     -6.6      -6.2 -6.6   0
## age.intercept36-47:rural -6.9 0.23       -7.3     -6.9      -6.4 -6.9   0
## age.intercept48-59:rural -7.3 0.29       -7.9     -7.3      -6.8 -7.3   0
## age.intercept0:urban     -2.7 0.13       -3.0     -2.7      -2.5 -2.7   0
## age.intercept1-11:urban  -5.0 0.13       -5.2     -5.0      -4.7 -5.0   0
## age.intercept12-23:urban -5.7 0.17       -6.1     -5.7      -5.4 -5.7   0
## age.intercept24-35:urban -7.1 0.29       -7.6     -7.1      -6.5 -7.1   0
## age.intercept36-47:urban -7.6 0.37       -8.3     -7.6      -6.9 -7.6   0
## age.intercept48-59:urban -8.0 0.46       -8.9     -8.0      -7.1 -8.0   0
## 
## Slope fixed effect index:
## time.slope.group1: 0:rural
## time.slope.group2: 1-11:rural
## time.slope.group3: 12-23:rural, 24-35:rural, 36-47:rural, 48-59:rural
## time.slope.group4: 0:urban
## time.slope.group5: 1-11:urban
## time.slope.group6: 12-23:urban, 24-35:urban, 36-47:urban, 48-59:urban
## ----------------------------------
## Random Effects
##            Name             Model
## 1   time.struct         RW2 model
## 2 time.unstruct         IID model
## 3 region.struct        BYM2 model
## 4    region.int Besags ICAR model
## 5   st.slope.id         IID model
## 6     survey.id         IID model
## ----------------------------------
## Model hyperparameters
##                                                     mean       sd 0.025quant 0.5quant
## overdispersion for the betabinomial observations   0.002    0.001      0.001    0.002
## Precision for time.struct                         91.454  111.521     11.674   58.458
## Precision for time.unstruct                      831.467 3076.794     15.769  225.650
## Precision for region.struct                      252.997  671.939      5.264   88.691
## Phi for region.struct                              0.346    0.240      0.027    0.297
## Precision for region.int                         771.901 3485.913      7.931  168.610
## Group PACF1 for region.int                         0.893    0.203      0.253    0.970
## Precision for st.slope.id                         24.590   63.442      0.801    9.168
##                                                  0.97quant   mode
## overdispersion for the betabinomial observations     0.005  0.002
## Precision for time.struct                          349.819 27.909
## Precision for time.unstruct                       4750.233 35.610
## Precision for region.struct                       1367.096 11.338
## Phi for region.struct                                0.849  0.077
## Precision for region.int                          4531.658 15.399
## Group PACF1 for region.int                           0.999  1.000
## Precision for st.slope.id                          130.531  1.920
##                                        [,1]
## log marginal-likelihood (integration) -3455
## log marginal-likelihood (Gaussian)    -3449
\end{verbatim}

The \texttt{est.bb} object above computes the U5MR estimates by urban/rural strata. In order to obtain the overall region-specific U5MR, we need additional information on the population fractions of each stratum within regions. For illustration purpose, here we simulate the population totals over the years and use these population totals to compute the population fractions later.

\begin{Shaded}
\begin{Highlighting}[]
\NormalTok{pop.base }\OtherTok{\textless{}{-}} \FunctionTok{expand.grid}\NormalTok{(}\AttributeTok{region =} \FunctionTok{c}\NormalTok{(}\StringTok{"central"}\NormalTok{, }\StringTok{"eastern"}\NormalTok{, }\StringTok{"northern"}\NormalTok{, }\StringTok{"western"}\NormalTok{), }
                   \AttributeTok{strata =} \FunctionTok{c}\NormalTok{(}\StringTok{"urban"}\NormalTok{, }\StringTok{"rural"}\NormalTok{))}
\NormalTok{pop.base}\SpecialCharTok{$}\NormalTok{population }\OtherTok{\textless{}{-}} \FunctionTok{round}\NormalTok{(}\FunctionTok{runif}\NormalTok{(}\FunctionTok{dim}\NormalTok{(pop.base)[}\DecValTok{1}\NormalTok{], }\DecValTok{1000}\NormalTok{, }\DecValTok{20000}\NormalTok{))}
\NormalTok{periods.all }\OtherTok{\textless{}{-}} \FunctionTok{c}\NormalTok{(periods, }\StringTok{"15{-}19"}\NormalTok{)}
\NormalTok{pop }\OtherTok{\textless{}{-}} \ConstantTok{NULL}
\ControlFlowTok{for}\NormalTok{(i }\ControlFlowTok{in} \DecValTok{1}\SpecialCharTok{:}\FunctionTok{length}\NormalTok{(periods.all))\{}
\NormalTok{  tmp }\OtherTok{\textless{}{-}}\NormalTok{ pop.base}
\NormalTok{  tmp}\SpecialCharTok{$}\NormalTok{population }\OtherTok{\textless{}{-}}\NormalTok{ pop.base}\SpecialCharTok{$}\NormalTok{population }\SpecialCharTok{+} \FunctionTok{round}\NormalTok{(}\FunctionTok{rnorm}\NormalTok{(}\FunctionTok{dim}\NormalTok{(pop.base)[}\DecValTok{1}\NormalTok{], }\AttributeTok{mean =} \DecValTok{0}\NormalTok{, }\AttributeTok{sd =} \DecValTok{200}\NormalTok{))}
\NormalTok{  tmp}\SpecialCharTok{$}\NormalTok{years }\OtherTok{\textless{}{-}}\NormalTok{ periods.all[i]}
\NormalTok{  pop }\OtherTok{\textless{}{-}} \FunctionTok{rbind}\NormalTok{(pop, tmp)}
\NormalTok{\}}
\FunctionTok{head}\NormalTok{(pop)}
\end{Highlighting}
\end{Shaded}

\begin{verbatim}
##     region strata population years
## 1  central  urban      16529 85-89
## 2  eastern  urban      10630 85-89
## 3 northern  urban      14664 85-89
## 4  western  urban      11470 85-89
## 5  central  rural       5836 85-89
## 6  eastern  rural       5479 85-89
\end{verbatim}

In order to compute the aggregated estimates, we need the proportion of urban/rural populations within each region in each time period, as computed below in the \texttt{weight.strata} object.

\begin{Shaded}
\begin{Highlighting}[]
\NormalTok{weight.strata }\OtherTok{\textless{}{-}} \FunctionTok{expand.grid}\NormalTok{(}\AttributeTok{region =} \FunctionTok{c}\NormalTok{(}\StringTok{"central"}\NormalTok{, }\StringTok{"eastern"}\NormalTok{, }\StringTok{"northern"}\NormalTok{, }\StringTok{"western"}\NormalTok{), }
                   \AttributeTok{years =}\NormalTok{ periods.all)}
\NormalTok{weight.strata}\SpecialCharTok{$}\NormalTok{urban }\OtherTok{\textless{}{-}}\NormalTok{ weight.strata}\SpecialCharTok{$}\NormalTok{rural }\OtherTok{\textless{}{-}} \ConstantTok{NA} 
\ControlFlowTok{for}\NormalTok{(i }\ControlFlowTok{in} \DecValTok{1}\SpecialCharTok{:}\FunctionTok{dim}\NormalTok{(weight.strata)[}\DecValTok{1}\NormalTok{])\{}
\NormalTok{  which.u }\OtherTok{\textless{}{-}} \FunctionTok{which}\NormalTok{(pop}\SpecialCharTok{$}\NormalTok{region }\SpecialCharTok{==}\NormalTok{ weight.strata}\SpecialCharTok{$}\NormalTok{region[i] }\SpecialCharTok{\&} 
\NormalTok{                   pop}\SpecialCharTok{$}\NormalTok{years }\SpecialCharTok{==}\NormalTok{ weight.strata}\SpecialCharTok{$}\NormalTok{years[i] }\SpecialCharTok{\&} 
\NormalTok{                   pop}\SpecialCharTok{$}\NormalTok{strata }\SpecialCharTok{==} \StringTok{"urban"}\NormalTok{)}
\NormalTok{  which.r }\OtherTok{\textless{}{-}} \FunctionTok{which}\NormalTok{(pop}\SpecialCharTok{$}\NormalTok{region }\SpecialCharTok{==}\NormalTok{ weight.strata}\SpecialCharTok{$}\NormalTok{region[i] }\SpecialCharTok{\&} 
\NormalTok{                   pop}\SpecialCharTok{$}\NormalTok{years }\SpecialCharTok{==}\NormalTok{ weight.strata}\SpecialCharTok{$}\NormalTok{years[i] }\SpecialCharTok{\&} 
\NormalTok{                   pop}\SpecialCharTok{$}\NormalTok{strata }\SpecialCharTok{==} \StringTok{"rural"}\NormalTok{)}
\NormalTok{  weight.strata[i, }\StringTok{"urban"}\NormalTok{] }\OtherTok{\textless{}{-}}\NormalTok{ pop}\SpecialCharTok{$}\NormalTok{population[which.u] }\SpecialCharTok{/} 
\NormalTok{                        (pop}\SpecialCharTok{$}\NormalTok{population[which.u] }\SpecialCharTok{+}\NormalTok{ pop}\SpecialCharTok{$}\NormalTok{population[which.r])}
\NormalTok{  weight.strata[i, }\StringTok{"rural"}\NormalTok{] }\OtherTok{\textless{}{-}} \DecValTok{1} \SpecialCharTok{{-}}\NormalTok{ weight.strata[i, }\StringTok{"urban"}\NormalTok{]}
\NormalTok{\}}
\FunctionTok{head}\NormalTok{(weight.strata)}
\end{Highlighting}
\end{Shaded}

\begin{verbatim}
##     region years rural urban
## 1  central 85-89  0.26  0.74
## 2  eastern 85-89  0.34  0.66
## 3 northern 85-89  0.53  0.47
## 4  western 85-89  0.34  0.66
## 5  central 90-94  0.26  0.74
## 6  eastern 90-94  0.36  0.64
\end{verbatim}

Now we can recompute the smoothed estimates with the population fractions.

\begin{Shaded}
\begin{Highlighting}[]
\NormalTok{est.bb }\OtherTok{\textless{}{-}} \FunctionTok{getSmoothed}\NormalTok{(fit.bb, }\AttributeTok{nsim =} \DecValTok{1000}\NormalTok{, }\AttributeTok{CI =} \FloatTok{0.95}\NormalTok{, }\AttributeTok{save.draws =} \ConstantTok{TRUE}\NormalTok{, }
                      \AttributeTok{weight.strata =}\NormalTok{ weight.strata)}
\FunctionTok{head}\NormalTok{(est.bb}\SpecialCharTok{$}\NormalTok{overall)}
\end{Highlighting}
\end{Shaded}

\begin{verbatim}
##    region years time area variance median mean upper lower rural urban is.yearly
## 5 central 85-89    1    1  0.00149   0.22 0.22  0.30  0.16  0.26  0.74     FALSE
## 6 central 90-94    2    1  0.00060   0.20 0.20  0.25  0.15  0.26  0.74     FALSE
## 7 central 95-99    3    1  0.00033   0.18 0.18  0.22  0.15  0.25  0.75     FALSE
## 1 central 00-04    4    1  0.00026   0.17 0.17  0.21  0.14  0.25  0.75     FALSE
## 2 central 05-09    5    1  0.00027   0.16 0.16  0.20  0.13  0.26  0.74     FALSE
## 3 central 10-14    6    1  0.00043   0.15 0.15  0.19  0.11  0.26  0.74     FALSE
##   years.num
## 5        NA
## 6        NA
## 7        NA
## 1        NA
## 2        NA
## 3        NA
\end{verbatim}

We can compare the stratum-specific and aggregated U5MR estimates now.

\begin{Shaded}
\begin{Highlighting}[]
\NormalTok{g1 }\OtherTok{\textless{}{-}} \FunctionTok{plot}\NormalTok{(est.bb}\SpecialCharTok{$}\NormalTok{stratified, }\AttributeTok{plot.CI =} \ConstantTok{TRUE}\NormalTok{) }\SpecialCharTok{+} \FunctionTok{facet\_wrap}\NormalTok{(}\SpecialCharTok{\textasciitilde{}}\NormalTok{strata) }\SpecialCharTok{+} \FunctionTok{ylim}\NormalTok{(}\DecValTok{0}\NormalTok{, }\FloatTok{0.5}\NormalTok{)}
\NormalTok{g2 }\OtherTok{\textless{}{-}} \FunctionTok{plot}\NormalTok{(est.bb}\SpecialCharTok{$}\NormalTok{overall, }\AttributeTok{plot.CI =} \ConstantTok{TRUE}\NormalTok{)  }\SpecialCharTok{+} \FunctionTok{ylim}\NormalTok{(}\DecValTok{0}\NormalTok{, }\FloatTok{0.5}\NormalTok{) }\SpecialCharTok{+} \FunctionTok{ggtitle}\NormalTok{(}\StringTok{"Aggregated estimates"}\NormalTok{)}
\NormalTok{g1 }\SpecialCharTok{+}\NormalTok{ g2}
\end{Highlighting}
\end{Shaded}

\begin{figure}[!ht]
\includegraphics[width=1\linewidth,]{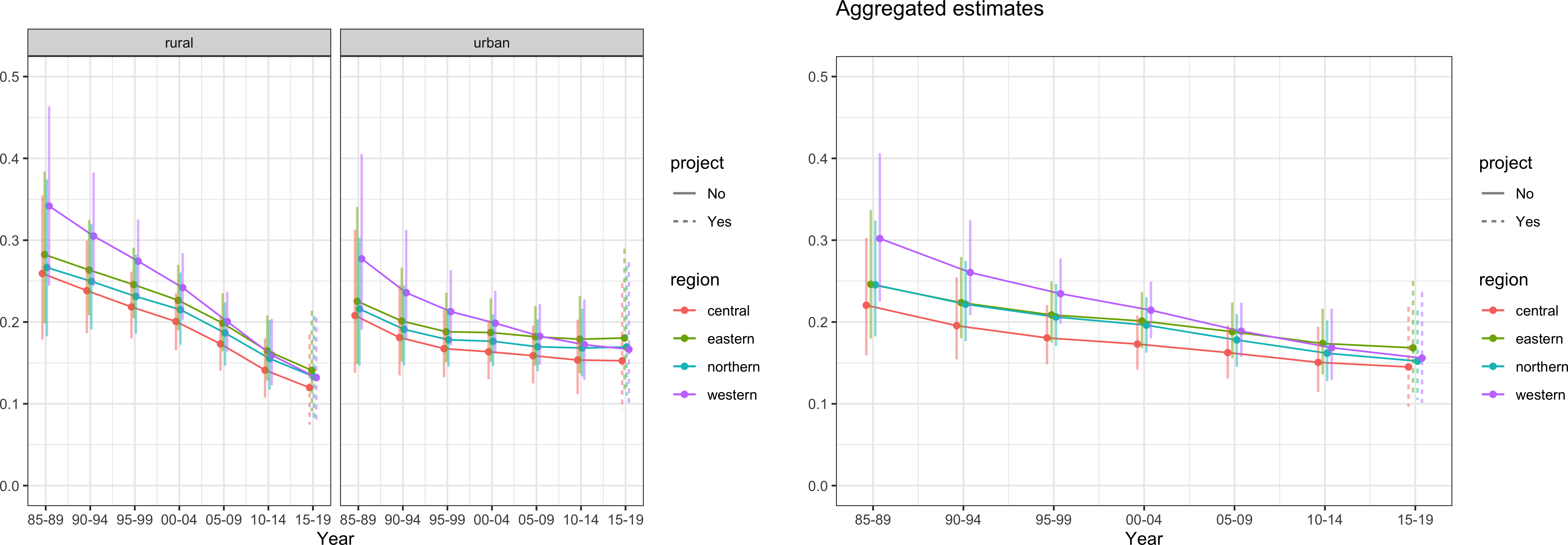} \caption{Comparing stratum-specific estimates of U5MR and the aggregated U5MR using simulated population weights.}\label{fig:sim-fit-s-6}
\end{figure}

\section{Obtaining and pre-processing Malawi DHS data}

In this section, we present the full workflow of downloading and pre-processing the two most recent Malawi DHS datasets, the 2010 and 2015 -- 2016 DHS. We also describe how the count data used in the main paper is produced. First, we load and process the spatial polygon data.

\begin{Shaded}
\begin{Highlighting}[]
\FunctionTok{data}\NormalTok{(MalawiMap)}
\NormalTok{MalawiGraph }\OtherTok{\textless{}{-}} \FunctionTok{getAmat}\NormalTok{(MalawiMap, }\AttributeTok{names=}\NormalTok{MalawiMap}\SpecialCharTok{$}\NormalTok{ADM2\_EN)}
\FunctionTok{mapPlot}\NormalTok{(}\AttributeTok{geo=}\NormalTok{MalawiMap, }\AttributeTok{by.geo =} \StringTok{"ADM2\_EN"}\NormalTok{)}
\end{Highlighting}
\end{Shaded}

\begin{figure}[!ht]

{\centering \includegraphics[width=.3\textwidth,]{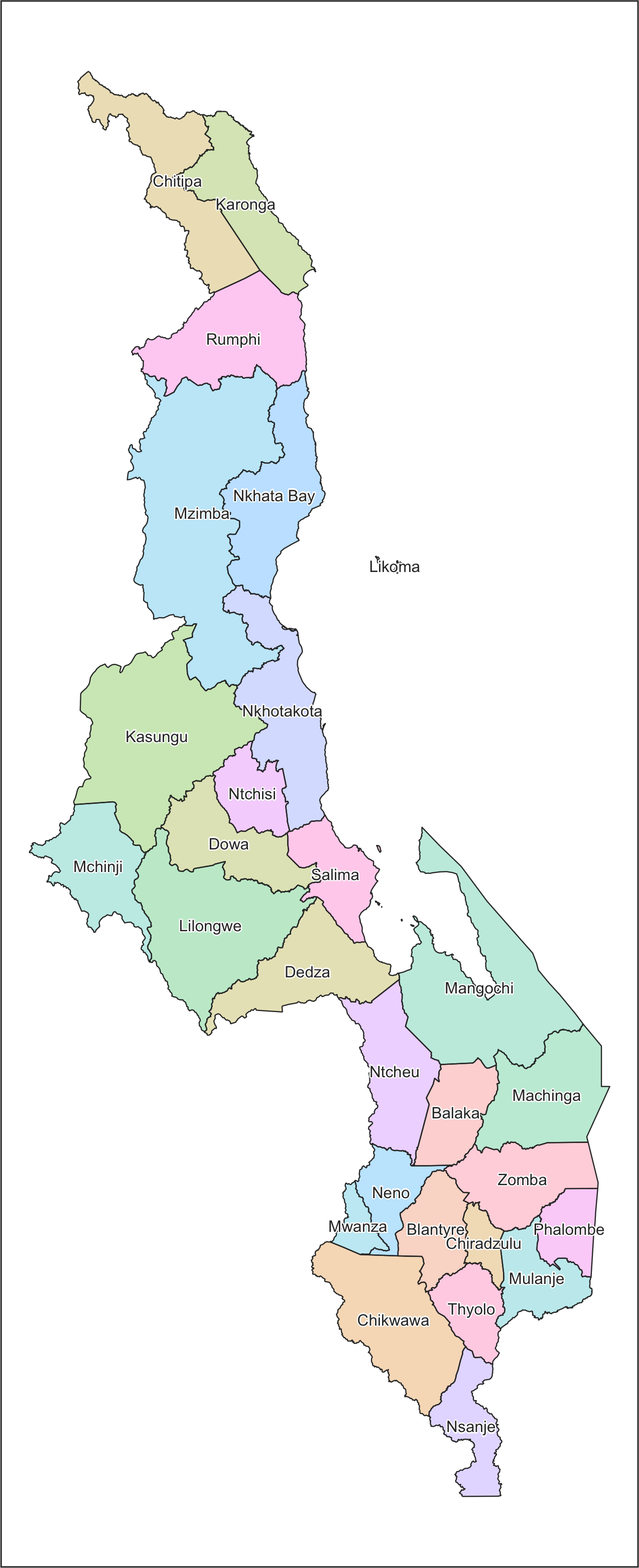} 

}

\caption{Admin-2 regions in Malawi.}\label{fig:load-map-1}
\end{figure}

\subsection{Loading DHS data using the \pkg{rdhs} package}

In this example, we use the 2010 and 2015--2016 Malawi DHS surveys. The DHS website (\url{https://dhsprogram.com/data/available-datasets.cfm?ctryid=24}) provides links to download all the surveys with registration. Once access is approved, we can use the \pkg{rdhs} package to load data directly from the DHS API \citep{rdhs}. We first download both the birth records (BR) and GPS data (GE) for these two surveys. Notice the following codes can only be run after registration with DHS and set up the credentials using the \pkg{rdhs} package. Additional information on this step can be found in the documentation of the \pkg{rdhs} package.

\begin{Shaded}
\begin{Highlighting}[]
\FunctionTok{library}\NormalTok{(rdhs)}
\NormalTok{sv }\OtherTok{\textless{}{-}} \FunctionTok{dhs\_surveys}\NormalTok{(}\AttributeTok{countryIds =} \StringTok{"MW"}\NormalTok{, }\AttributeTok{surveyType =} \StringTok{"DHS"}\NormalTok{, }
                  \AttributeTok{surveyYearStart =} \DecValTok{2010}\NormalTok{)}
\NormalTok{BR }\OtherTok{\textless{}{-}} \FunctionTok{dhs\_datasets}\NormalTok{(}\AttributeTok{surveyIds =}\NormalTok{ sv}\SpecialCharTok{$}\NormalTok{SurveyId, }\AttributeTok{fileFormat =} \StringTok{"Flat"}\NormalTok{, }
                   \AttributeTok{fileType =} \StringTok{"BR"}\NormalTok{)}
\NormalTok{BRfiles }\OtherTok{\textless{}{-}} \FunctionTok{get\_datasets}\NormalTok{(BR}\SpecialCharTok{$}\NormalTok{FileName, }\AttributeTok{reformat=}\ConstantTok{TRUE}\NormalTok{)}
\NormalTok{GPS }\OtherTok{\textless{}{-}} \FunctionTok{dhs\_datasets}\NormalTok{(}\AttributeTok{surveyIds =}\NormalTok{ sv}\SpecialCharTok{$}\NormalTok{SurveyId, }\AttributeTok{fileFormat =} \StringTok{"Flat"}\NormalTok{, }
                    \AttributeTok{fileType =} \StringTok{"GE"}\NormalTok{)}
\NormalTok{GPSfiles }\OtherTok{\textless{}{-}} \FunctionTok{get\_datasets}\NormalTok{(GPS}\SpecialCharTok{$}\NormalTok{FileName, }\AttributeTok{reformat=}\ConstantTok{TRUE}\NormalTok{)}
\end{Highlighting}
\end{Shaded}

The downloaded data can then be loaded by the returned file paths.

\begin{Shaded}
\begin{Highlighting}[]
\NormalTok{Surv2010 }\OtherTok{\textless{}{-}} \FunctionTok{readRDS}\NormalTok{(BRfiles[[}\DecValTok{1}\NormalTok{]])}
\NormalTok{Surv2015 }\OtherTok{\textless{}{-}} \FunctionTok{readRDS}\NormalTok{(BRfiles[[}\DecValTok{2}\NormalTok{]])}
\NormalTok{DHS2010.geo }\OtherTok{\textless{}{-}} \FunctionTok{readRDS}\NormalTok{(GPSfiles[[}\DecValTok{1}\NormalTok{]])}
\NormalTok{DHS2015.geo }\OtherTok{\textless{}{-}} \FunctionTok{readRDS}\NormalTok{(GPSfiles[[}\DecValTok{2}\NormalTok{]])}
\end{Highlighting}
\end{Shaded}

\begin{Shaded}
\begin{Highlighting}[]
\FunctionTok{load}\NormalTok{(}\StringTok{"data/downloaded.rda"}\NormalTok{)}
\end{Highlighting}
\end{Shaded}

\subsection{Cleaning the DHS data}

We then use the \texttt{getBirths()} function to process the raw birth history into person-month format. We label the person-month records with 6 age groups specified by \texttt{month.cut}. In this example, instead of using 5 year periods, we work directly with yearly estimates specified by \texttt{year.cut}.

\begin{Shaded}
\begin{Highlighting}[]
\NormalTok{DHS2010 }\OtherTok{\textless{}{-}} \FunctionTok{getBirths}\NormalTok{(}\AttributeTok{data =}\NormalTok{ Surv2010,}
                  \AttributeTok{month.cut =} \FunctionTok{c}\NormalTok{(}\DecValTok{1}\NormalTok{, }\DecValTok{12}\NormalTok{, }\DecValTok{24}\NormalTok{, }\DecValTok{36}\NormalTok{, }\DecValTok{48}\NormalTok{, }\DecValTok{60}\NormalTok{),}
                  \AttributeTok{year.cut =} \FunctionTok{seq}\NormalTok{(}\DecValTok{2000}\NormalTok{, }\DecValTok{2020}\NormalTok{, }\AttributeTok{by =} \DecValTok{1}\NormalTok{), }\AttributeTok{strata =} \StringTok{"v022"}\NormalTok{)}
\NormalTok{DHS2015 }\OtherTok{\textless{}{-}} \FunctionTok{getBirths}\NormalTok{(}\AttributeTok{data =}\NormalTok{ Surv2015,}
                  \AttributeTok{month.cut =} \FunctionTok{c}\NormalTok{(}\DecValTok{1}\NormalTok{, }\DecValTok{12}\NormalTok{, }\DecValTok{24}\NormalTok{, }\DecValTok{36}\NormalTok{, }\DecValTok{48}\NormalTok{, }\DecValTok{60}\NormalTok{),}
                  \AttributeTok{year.cut =} \FunctionTok{seq}\NormalTok{(}\DecValTok{2000}\NormalTok{, }\DecValTok{2020}\NormalTok{, }\AttributeTok{by =} \DecValTok{1}\NormalTok{), }\AttributeTok{strata =} \StringTok{"v022"}\NormalTok{)}
\end{Highlighting}
\end{Shaded}

We perform similar processing steps for the 2015--2016 DHS survey birth records. Since only a small fraction of observations are available in 2016, we remove the partial year observations in this survey.

\begin{Shaded}
\begin{Highlighting}[]
\NormalTok{DHS2015 }\OtherTok{\textless{}{-}} \FunctionTok{subset}\NormalTok{(DHS2015, time }\SpecialCharTok{!=} \DecValTok{2016}\NormalTok{)}
\end{Highlighting}
\end{Shaded}

The DHS dataset does not usually have sufficient spatial resolution information in the birth records file, so we need to use the GPS datasets of cluster locations to assign records to the admin-2 areas. This process can be highly survey-specific and require more extensive data manipulation. In the 2010 DHS dataset, the GPS file contains the cluster ID (\texttt{DHSCLUST}), urbanicity indicator (\texttt{URBAN\_RURA}), and \texttt{DHSCLUST} variable in the format of ``admin-2\_urbanrural'', with some spelling and capitalization differences as in the polygon file. From the GPS dataset, we processed a list of clusters and their corresponding admin-2 areas.

\begin{Shaded}
\begin{Highlighting}[]
\NormalTok{cluster.list }\OtherTok{\textless{}{-}} \FunctionTok{data.frame}\NormalTok{(DHS2010.geo) }\SpecialCharTok{\%\textgreater{}\%} 
         \FunctionTok{distinct}\NormalTok{(DHSCLUST, DHSREGNA, URBAN\_RURA) }\SpecialCharTok{\%\textgreater{}\%}
         \FunctionTok{mutate}\NormalTok{(}\AttributeTok{admin2 =} \FunctionTok{gsub}\NormalTok{(}\StringTok{" {-} rural"}\NormalTok{, }\StringTok{""}\NormalTok{, DHSREGNA)) }\SpecialCharTok{\%\textgreater{}\%} 
         \FunctionTok{mutate}\NormalTok{(}\AttributeTok{admin2 =} \FunctionTok{gsub}\NormalTok{(}\StringTok{" {-} urban"}\NormalTok{, }\StringTok{""}\NormalTok{, admin2)) }\SpecialCharTok{\%\textgreater{}\%}
         \FunctionTok{mutate}\NormalTok{(}\AttributeTok{admin2 =} \FunctionTok{recode}\NormalTok{(admin2, }\StringTok{"nkhatabay"} \OtherTok{=} \StringTok{"nkhata bay"}\NormalTok{)) }\SpecialCharTok{\%\textgreater{}\%}
         \FunctionTok{mutate}\NormalTok{(}\AttributeTok{admin2 =} \FunctionTok{recode}\NormalTok{(admin2, }\StringTok{"nkhota kota"} \OtherTok{=} \StringTok{"nkhotakota"}\NormalTok{)) }\SpecialCharTok{\%\textgreater{}\%}
         \FunctionTok{mutate}\NormalTok{(}\AttributeTok{admin2 =}  \FunctionTok{str\_to\_title}\NormalTok{(admin2)) }\SpecialCharTok{\%\textgreater{}\%}
         \FunctionTok{mutate}\NormalTok{(}\AttributeTok{urban =} \FunctionTok{ifelse}\NormalTok{(URBAN\_RURA}\SpecialCharTok{==}\StringTok{"U"}\NormalTok{, }\DecValTok{1}\NormalTok{, }\DecValTok{0}\NormalTok{))  }\SpecialCharTok{\%\textgreater{}\%}
         \FunctionTok{select}\NormalTok{(}\AttributeTok{v001 =}\NormalTok{ DHSCLUST, admin2, urban) }
\FunctionTok{head}\NormalTok{(cluster.list, }\AttributeTok{n=}\DecValTok{3}\NormalTok{)     }
\end{Highlighting}
\end{Shaded}

\begin{verbatim}
##   v001  admin2 urban
## 1    1   Dedza     0
## 2    2  Balaka     0
## 3    3 Mchinji     0
\end{verbatim}

We then merge this list to the main person-month file, which adds the \texttt{admin2} and \texttt{urban} columns to the data frame.

\begin{Shaded}
\begin{Highlighting}[]
\NormalTok{DHS2010 }\OtherTok{\textless{}{-}}\NormalTok{ DHS2010 }\SpecialCharTok{\%\textgreater{}\%} \FunctionTok{left\_join}\NormalTok{(cluster.list)}
\FunctionTok{head}\NormalTok{(DHS2010)}
\end{Highlighting}
\end{Shaded}

\begin{verbatim}
##    dob survey_year died id.new          caseid v001 v002 v004    v005 v021 v022    v023
## 1 1316          NA    0      1         1  1  2    1    1    1 1902867    1   12 central
## 2 1316          NA    0      1         1  1  2    1    1    1 1902867    1   12 central
## 3 1316          NA    0      1         1  1  2    1    1    1 1902867    1   12 central
## 4 1316          NA    0      1         1  1  2    1    1    1 1902867    1   12 central
## 5 1316          NA    0      1         1  1  2    1    1    1 1902867    1   12 central
## 6 1316          NA    0      1         1  1  2    1    1    1 1902867    1   12 central
##      v024  v025    v139 bidx agemonth obsStart obsStop obsmonth year  age time strata
## 1 central rural central    1        0     1316    1316     1316  109    0 2009     12
## 2 central rural central    1        1     1317    1317     1317  109 1-11 2009     12
## 3 central rural central    1        2     1318    1318     1318  109 1-11 2009     12
## 4 central rural central    1        3     1319    1319     1319  109 1-11 2009     12
## 5 central rural central    1        4     1320    1320     1320  109 1-11 2009     12
## 6 central rural central    1        5     1321    1321     1321  110 1-11 2010     12
##   admin2 urban
## 1  Dedza     0
## 2  Dedza     0
## 3  Dedza     0
## 4  Dedza     0
## 5  Dedza     0
## 6  Dedza     0
\end{verbatim}

We perform similar processing steps for the 2015--2016 DHS GPS data. To illustrate the idiosyncratic data processing steps required for each survey, the \texttt{DHSCLUST} variable in this dataset is in the form of ``admin-1\_urbanrural'' and does not allow us to parse the admin-2 area names. The strata variable \texttt{v022} in the birth records, however, is in the ``admin-2\_urbanrural'' format. Therefore, we first merge the \texttt{v022} variables to the GPS data and proceed in a similar fashion to the previous example.

\begin{Shaded}
\begin{Highlighting}[]
\NormalTok{cluster.list }\OtherTok{\textless{}{-}} \FunctionTok{data.frame}\NormalTok{(DHS2015.geo) }\SpecialCharTok{\%\textgreater{}\%} 
         \FunctionTok{mutate}\NormalTok{(}\AttributeTok{v001 =}\NormalTok{ DHSCLUST) }\SpecialCharTok{\%\textgreater{}\%}
         \FunctionTok{left\_join}\NormalTok{(DHS2015, }\AttributeTok{by =} \StringTok{"v001"}\NormalTok{) }\SpecialCharTok{\%\textgreater{}\%} 
         \FunctionTok{distinct}\NormalTok{(v001, v022, URBAN\_RURA) }\SpecialCharTok{\%\textgreater{}\%}
         \FunctionTok{mutate}\NormalTok{(}\AttributeTok{admin2 =} \FunctionTok{gsub}\NormalTok{(}\StringTok{" {-} rural"}\NormalTok{, }\StringTok{""}\NormalTok{, v022)) }\SpecialCharTok{\%\textgreater{}\%} 
         \FunctionTok{mutate}\NormalTok{(}\AttributeTok{admin2 =} \FunctionTok{gsub}\NormalTok{(}\StringTok{" {-} urban"}\NormalTok{, }\StringTok{""}\NormalTok{, admin2)) }\SpecialCharTok{\%\textgreater{}\%}
         \FunctionTok{mutate}\NormalTok{(}\AttributeTok{admin2 =} \FunctionTok{recode}\NormalTok{(admin2, }\StringTok{"nkhatabay"} \OtherTok{=} \StringTok{"nkhata bay"}\NormalTok{)) }\SpecialCharTok{\%\textgreater{}\%}
         \FunctionTok{mutate}\NormalTok{(}\AttributeTok{admin2 =} \FunctionTok{recode}\NormalTok{(admin2, }\StringTok{"nkhota kota"} \OtherTok{=} \StringTok{"nkhotakota"}\NormalTok{)) }\SpecialCharTok{\%\textgreater{}\%}
         \FunctionTok{mutate}\NormalTok{(}\AttributeTok{admin2 =}  \FunctionTok{str\_to\_title}\NormalTok{(admin2)) }\SpecialCharTok{\%\textgreater{}\%}
         \FunctionTok{mutate}\NormalTok{(}\AttributeTok{urban =} \FunctionTok{ifelse}\NormalTok{(URBAN\_RURA}\SpecialCharTok{==}\StringTok{"U"}\NormalTok{, }\DecValTok{1}\NormalTok{, }\DecValTok{0}\NormalTok{))  }\SpecialCharTok{\%\textgreater{}\%}
         \FunctionTok{select}\NormalTok{(v001, admin2, urban) }
\NormalTok{DHS2015 }\OtherTok{\textless{}{-}}\NormalTok{ DHS2015 }\SpecialCharTok{\%\textgreater{}\%} \FunctionTok{left\_join}\NormalTok{(cluster.list)}
\FunctionTok{head}\NormalTok{(DHS2015)}
\end{Highlighting}
\end{Shaded}

\begin{verbatim}
##    dob survey_year died id.new          caseid v001 v002 v004   v005 v021            v022
## 1 1334          NA    0      1        1  12  2    1   12    1 118748    1 ntchisi - urban
## 2 1334          NA    0      1        1  12  2    1   12    1 118748    1 ntchisi - urban
## 3 1334          NA    0      1        1  12  2    1   12    1 118748    1 ntchisi - urban
## 4 1334          NA    0      1        1  12  2    1   12    1 118748    1 ntchisi - urban
## 5 1334          NA    0      1        1  12  2    1   12    1 118748    1 ntchisi - urban
## 6 1334          NA    0      1        1  12  2    1   12    1 118748    1 ntchisi - urban
##              v023           v024  v025           v139 bidx agemonth obsStart obsStop
## 1 ntchisi - urban central region urban central region    1        0     1334    1334
## 2 ntchisi - urban central region urban central region    1        1     1335    1335
## 3 ntchisi - urban central region urban central region    1        2     1336    1336
## 4 ntchisi - urban central region urban central region    1        3     1337    1337
## 5 ntchisi - urban central region urban central region    1        4     1338    1338
## 6 ntchisi - urban central region urban central region    1        5     1339    1339
##   obsmonth year  age time          strata  admin2 urban
## 1     1334  111    0 2011 ntchisi - urban Ntchisi     1
## 2     1335  111 1-11 2011 ntchisi - urban Ntchisi     1
## 3     1336  111 1-11 2011 ntchisi - urban Ntchisi     1
## 4     1337  111 1-11 2011 ntchisi - urban Ntchisi     1
## 5     1338  111 1-11 2011 ntchisi - urban Ntchisi     1
## 6     1339  111 1-11 2011 ntchisi - urban Ntchisi     1
\end{verbatim}

Then we use the \texttt{getCounts()} function to obtain the count data format input and stack the two DHS survey data into a single data frame.

\begin{Shaded}
\begin{Highlighting}[]
\NormalTok{vars }\OtherTok{\textless{}{-}} \FunctionTok{c}\NormalTok{(}\StringTok{"v001"}\NormalTok{, }\StringTok{"v025"}\NormalTok{, }\StringTok{"admin2"}\NormalTok{, }\StringTok{"time"}\NormalTok{, }\StringTok{"age"}\NormalTok{, }\StringTok{"v005"}\NormalTok{)}
\NormalTok{dat1 }\OtherTok{\textless{}{-}} \FunctionTok{getCounts}\NormalTok{(DHS2010[, }\FunctionTok{c}\NormalTok{(vars, }\StringTok{"died"}\NormalTok{)], }\AttributeTok{variables =} \StringTok{\textquotesingle{}died\textquotesingle{}}\NormalTok{, }
                  \AttributeTok{by =}\NormalTok{ vars, }\AttributeTok{drop=}\ConstantTok{TRUE}\NormalTok{)}
\NormalTok{dat1}\SpecialCharTok{$}\NormalTok{survey }\OtherTok{=} \StringTok{"DHS2010"}
\NormalTok{dat2 }\OtherTok{\textless{}{-}} \FunctionTok{getCounts}\NormalTok{(DHS2015[, }\FunctionTok{c}\NormalTok{(vars, }\StringTok{"died"}\NormalTok{)], }\AttributeTok{variables =} \StringTok{\textquotesingle{}died\textquotesingle{}}\NormalTok{, }
                  \AttributeTok{by =}\NormalTok{ vars, }\AttributeTok{drop=}\ConstantTok{TRUE}\NormalTok{)}
\NormalTok{dat2}\SpecialCharTok{$}\NormalTok{survey }\OtherTok{=} \StringTok{"DHS2015"}
\NormalTok{DHS.counts }\OtherTok{\textless{}{-}} \FunctionTok{rbind}\NormalTok{(dat1, dat2) }\SpecialCharTok{\%\textgreater{}\%} 
                \FunctionTok{mutate}\NormalTok{(}\AttributeTok{cluster =}\NormalTok{ v001, }\AttributeTok{strata =}\NormalTok{ v025, }\AttributeTok{region =}\NormalTok{ admin2, }
                       \AttributeTok{years =}\NormalTok{ time, }\AttributeTok{Y =}\NormalTok{ died)}
\end{Highlighting}
\end{Shaded}

The analysis in the main paper is carried out only on the 2015--2016 dataset, which is obtained using the same procedure.

\section{Additional analysis of the 2010 and 2015 -- 2016 Malawi DHS}

In this section, we provide additional workflow of estimating U5MR from 2000 to 2019 using the two DHS surveys. The focus of this section is on the models not discussed in details in Example 3 of the main paper.

\subsection{Direct estimates}

We use \texttt{getDirectList} to obtain direct estimates from both surveys and combine into the `meta-analysis' estimator using \texttt{aggregateSurvey}. Notice that the \texttt{survey} variable in the returned data frame contains a numeric index of each survey in the list, and the \texttt{surveyYears} contains the survey names we assigned in the input list.

\begin{Shaded}
\begin{Highlighting}[]
\NormalTok{direct }\OtherTok{\textless{}{-}} \FunctionTok{getDirectList}\NormalTok{(}\AttributeTok{births =} \FunctionTok{list}\NormalTok{(}\AttributeTok{DHS2010=}\NormalTok{DHS2010, }\AttributeTok{DHS2015=}\NormalTok{DHS2015), }
        \AttributeTok{years =} \DecValTok{2000}\SpecialCharTok{:}\DecValTok{2019}\NormalTok{, }\AttributeTok{regionVar =} \StringTok{"admin2"}\NormalTok{, }\AttributeTok{timeVar =} \StringTok{"time"}\NormalTok{, }
        \AttributeTok{clusterVar =} \StringTok{"\textasciitilde{}v001 + v002"}\NormalTok{, }\AttributeTok{ageVar =} \StringTok{"age"}\NormalTok{, }\AttributeTok{weightsVar =} \StringTok{"v005"}\NormalTok{)}
\NormalTok{direct.comb }\OtherTok{\textless{}{-}} \FunctionTok{aggregateSurvey}\NormalTok{(direct)}
\end{Highlighting}
\end{Shaded}

When additional information is available to adjust the direct estimates from the surveys, we use the methods described in \citet{li_etal_19}. We can perform the ratio adjustment to the direct estimates using the \texttt{getAdjusted()} function. For the two surveys in Malawi, the calculated HIV adjustment ratios as described in \citet{walker_etal_12} are stored in \texttt{MalawiData\$HIV.yearly}. In order to get the correct uncertainty bounds, we also need to specify the columns corresponding to the unadjusted uncertainty bounds, the \texttt{lower} and \texttt{upper} columns, which are on the probability scale in this case.

\begin{Shaded}
\begin{Highlighting}[]
\FunctionTok{data}\NormalTok{(MalawiData)}
\NormalTok{direct}\FloatTok{.2010} \OtherTok{\textless{}{-}} \FunctionTok{subset}\NormalTok{(direct, survey }\SpecialCharTok{==} \DecValTok{1}\NormalTok{)}
\NormalTok{direct.}\FloatTok{2010.}\NormalTok{hiv }\OtherTok{\textless{}{-}} \FunctionTok{getAdjusted}\NormalTok{(}\AttributeTok{data =}\NormalTok{ direct}\FloatTok{.2010}\NormalTok{, }
                \AttributeTok{ratio =} \FunctionTok{subset}\NormalTok{(MalawiData}\SpecialCharTok{$}\NormalTok{HIV.yearly, survey }\SpecialCharTok{==} \StringTok{"DHS2010"}\NormalTok{), }
                \AttributeTok{logit.lower =} \ConstantTok{NA}\NormalTok{, }\AttributeTok{logit.upper =} \ConstantTok{NA}\NormalTok{, }
                \AttributeTok{prob.lower =} \StringTok{"lower"}\NormalTok{, }\AttributeTok{prob.upper =} \StringTok{"upper"}\NormalTok{)}
\NormalTok{direct}\FloatTok{.2015} \OtherTok{\textless{}{-}} \FunctionTok{subset}\NormalTok{(direct, survey }\SpecialCharTok{==} \DecValTok{2}\NormalTok{)}
\NormalTok{direct.}\FloatTok{2015.}\NormalTok{hiv }\OtherTok{\textless{}{-}} \FunctionTok{getAdjusted}\NormalTok{(}\AttributeTok{data =}\NormalTok{ direct}\FloatTok{.2015}\NormalTok{, }
                \AttributeTok{ratio =} \FunctionTok{subset}\NormalTok{(MalawiData}\SpecialCharTok{$}\NormalTok{HIV.yearly, survey }\SpecialCharTok{==} \StringTok{"DHS2015"}\NormalTok{), }
                \AttributeTok{logit.lower =} \ConstantTok{NA}\NormalTok{, }\AttributeTok{logit.upper =} \ConstantTok{NA}\NormalTok{, }
                \AttributeTok{prob.lower =} \StringTok{"lower"}\NormalTok{, }\AttributeTok{prob.upper =} \StringTok{"upper"}\NormalTok{)}
\end{Highlighting}
\end{Shaded}

Finally, we combine the direct estimates into \texttt{direct.comb.hiv}.

\begin{Shaded}
\begin{Highlighting}[]
\NormalTok{direct.hiv }\OtherTok{\textless{}{-}} \FunctionTok{rbind}\NormalTok{(direct.}\FloatTok{2010.}\NormalTok{hiv, direct.}\FloatTok{2015.}\NormalTok{hiv)}
\NormalTok{direct.comb.hiv }\OtherTok{\textless{}{-}} \FunctionTok{aggregateSurvey}\NormalTok{(direct.hiv)}
\end{Highlighting}
\end{Shaded}

\subsection{Space-time Fay-Harriot estimates}

We now fit a national Fay-Harriot model with the calculated direct estimates using \texttt{smoothDirect()} and \texttt{getSmoothed()} functions.

\begin{Shaded}
\begin{Highlighting}[]
\NormalTok{fit.national.unadj }\OtherTok{\textless{}{-}} \FunctionTok{smoothDirect}\NormalTok{(}\AttributeTok{data =}\NormalTok{ direct.comb, }\AttributeTok{Amat =} \ConstantTok{NULL}\NormalTok{, }
                        \AttributeTok{year.label =} \DecValTok{2000}\SpecialCharTok{:}\DecValTok{2019}\NormalTok{, }\AttributeTok{year.range =} \FunctionTok{c}\NormalTok{(}\DecValTok{2000}\NormalTok{, }\DecValTok{2019}\NormalTok{), }
                        \AttributeTok{time.model =} \StringTok{"rw2"}\NormalTok{, }\AttributeTok{m =} \DecValTok{1}\NormalTok{)}
\NormalTok{est.unadj }\OtherTok{\textless{}{-}} \FunctionTok{getSmoothed}\NormalTok{(fit.national.unadj)}
\end{Highlighting}
\end{Shaded}

For comparison, we smooth both the unadjusted direct estimates and the direct estimates with HIV adjustments.

\begin{Shaded}
\begin{Highlighting}[]
\NormalTok{fit.national.hiv }\OtherTok{\textless{}{-}} \FunctionTok{smoothDirect}\NormalTok{(}\AttributeTok{data =}\NormalTok{ direct.comb.hiv,  }\AttributeTok{Amat =} \ConstantTok{NULL}\NormalTok{, }
                        \AttributeTok{year.label =} \DecValTok{2000}\SpecialCharTok{:}\DecValTok{2019}\NormalTok{, }\AttributeTok{year.range =} \FunctionTok{c}\NormalTok{(}\DecValTok{2000}\NormalTok{, }\DecValTok{2019}\NormalTok{), }
                        \AttributeTok{time.model =} \StringTok{"rw2"}\NormalTok{, }\AttributeTok{m =} \DecValTok{1}\NormalTok{)}
\NormalTok{est.hiv }\OtherTok{\textless{}{-}} \FunctionTok{getSmoothed}\NormalTok{(fit.national.hiv)}
\end{Highlighting}
\end{Shaded}

In addition, we also demonstrate the benchmarking procedure described in \citet{li_etal_19}, where we first fit a smoothing model and then benchmark the results with the UN IGME estimates -- the latter are based on more extensive data \citep{alkema_new_14} . We first calculate the adjustment ratio compared to the 2019 UN IGME estimates.

\begin{Shaded}
\begin{Highlighting}[]
\NormalTok{UN }\OtherTok{\textless{}{-}}\NormalTok{ MalawiData}\SpecialCharTok{$}\NormalTok{IGME2019}
\NormalTok{UN.est }\OtherTok{\textless{}{-}}\NormalTok{ UN}\SpecialCharTok{$}\NormalTok{mean[}\FunctionTok{match}\NormalTok{(}\DecValTok{2000}\SpecialCharTok{:}\DecValTok{2019}\NormalTok{, UN}\SpecialCharTok{$}\NormalTok{years)]}
\NormalTok{Smooth.est }\OtherTok{\textless{}{-}}\NormalTok{ est.hiv}\SpecialCharTok{$}\NormalTok{median[}\FunctionTok{match}\NormalTok{(}\DecValTok{2000}\SpecialCharTok{:}\DecValTok{2019}\NormalTok{, est.hiv}\SpecialCharTok{$}\NormalTok{years)]}
\NormalTok{UN.adj }\OtherTok{\textless{}{-}} \FunctionTok{data.frame}\NormalTok{(}\AttributeTok{years =} \DecValTok{2000}\SpecialCharTok{:}\DecValTok{2019}\NormalTok{, }\AttributeTok{ratio =}\NormalTok{ Smooth.est }\SpecialCharTok{/}\NormalTok{ UN.est)}
\FunctionTok{head}\NormalTok{(UN.adj, }\AttributeTok{n =} \DecValTok{3}\NormalTok{)}
\end{Highlighting}
\end{Shaded}

\begin{verbatim}
##   years ratio
## 1  2000   1.1
## 2  2001   1.2
## 3  2002   1.1
\end{verbatim}

We then fit the smoothing model on the benchmarked direct estimates.

\begin{Shaded}
\begin{Highlighting}[]
\NormalTok{direct.comb.benchmark }\OtherTok{\textless{}{-}} \FunctionTok{getAdjusted}\NormalTok{(}\AttributeTok{data =}\NormalTok{ direct.comb.hiv, }\AttributeTok{ratio =}\NormalTok{ UN.adj,}
                                \AttributeTok{logit.lower =} \ConstantTok{NA}\NormalTok{, }\AttributeTok{logit.upper =} \ConstantTok{NA}\NormalTok{, }
                                \AttributeTok{prob.lower =} \StringTok{"lower"}\NormalTok{, }\AttributeTok{prob.upper =} \StringTok{"upper"}\NormalTok{)}
\NormalTok{fit.benchmark }\OtherTok{\textless{}{-}} \FunctionTok{smoothDirect}\NormalTok{(}\AttributeTok{data =}\NormalTok{ direct.comb.benchmark, }\AttributeTok{Amat =} \ConstantTok{NULL}\NormalTok{, }
                         \AttributeTok{year.label =} \DecValTok{2000}\SpecialCharTok{:}\DecValTok{2019}\NormalTok{, }\AttributeTok{year.range =} \FunctionTok{c}\NormalTok{(}\DecValTok{2000}\NormalTok{, }\DecValTok{2019}\NormalTok{), }
                         \AttributeTok{time.model =} \StringTok{"rw2"}\NormalTok{,  }\AttributeTok{m =} \DecValTok{1}\NormalTok{)}
\NormalTok{est.benchmark }\OtherTok{\textless{}{-}} \FunctionTok{getSmoothed}\NormalTok{(fit.benchmark)}
\end{Highlighting}
\end{Shaded}

\begin{figure}[!ht]
\includegraphics[width=\textwidth,]{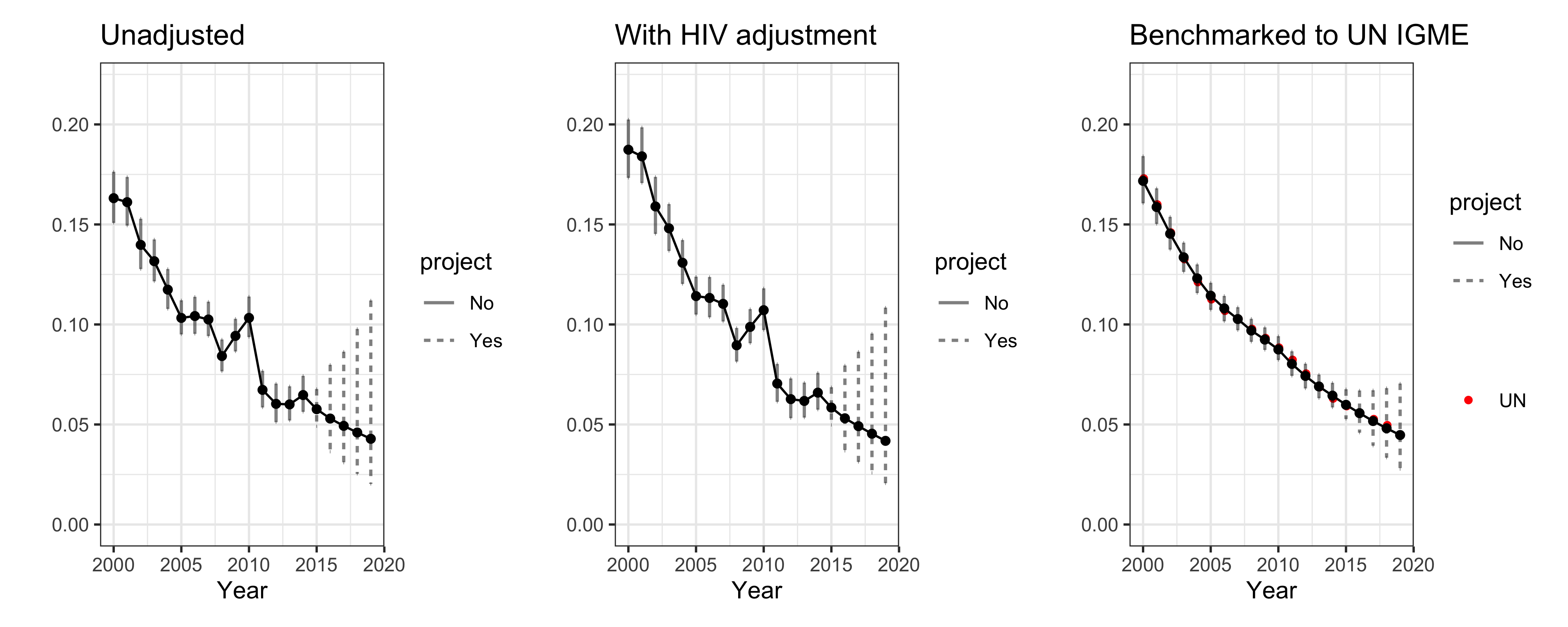} \caption{Comparison of the different Fay-Harriot estimates. The UN IGME estimates are plotted as red dots on the benchmarked plot.}\label{fig:national-5}
\end{figure}

We compare the different Fay-Harriot estimates in Figure \ref{fig:national-5}. Compared to the raw estimates, HIV adjustments lead to higher estimates in the earlier years. The benchmarking step produces a national trend that follows the same trajectory as the UN IGME estimates.

\begin{Shaded}
\begin{Highlighting}[]
\NormalTok{g1 }\OtherTok{\textless{}{-}} \FunctionTok{plot}\NormalTok{(est.unadj, }\AttributeTok{is.subnational=}\ConstantTok{FALSE}\NormalTok{, }\AttributeTok{proj.year =} \DecValTok{2016}\NormalTok{) }\SpecialCharTok{+} 
            \FunctionTok{ggtitle}\NormalTok{(}\StringTok{"Unadjusted"}\NormalTok{) }\SpecialCharTok{+} \FunctionTok{ylim}\NormalTok{(}\FunctionTok{c}\NormalTok{(}\DecValTok{0}\NormalTok{, }\FloatTok{0.22}\NormalTok{))}
\NormalTok{g2 }\OtherTok{\textless{}{-}} \FunctionTok{plot}\NormalTok{(est.hiv, }\AttributeTok{is.subnational=}\ConstantTok{FALSE}\NormalTok{, }\AttributeTok{proj.year =} \DecValTok{2016}\NormalTok{) }\SpecialCharTok{+} 
            \FunctionTok{ggtitle}\NormalTok{(}\StringTok{"With HIV adjustment"}\NormalTok{) }\SpecialCharTok{+} \FunctionTok{ylim}\NormalTok{(}\FunctionTok{c}\NormalTok{(}\DecValTok{0}\NormalTok{, }\FloatTok{0.22}\NormalTok{))}
\NormalTok{g3 }\OtherTok{\textless{}{-}} \FunctionTok{plot}\NormalTok{(est.benchmark, }\AttributeTok{is.subnational=}\ConstantTok{FALSE}\NormalTok{, }\AttributeTok{proj.year =} \DecValTok{2016}\NormalTok{, }\AttributeTok{data.add =}\NormalTok{ UN, }
    \AttributeTok{option.add =} \FunctionTok{list}\NormalTok{(}\AttributeTok{point =} \StringTok{"mean"}\NormalTok{), }\AttributeTok{label.add =} \StringTok{"UN"}\NormalTok{, }\AttributeTok{color.add =} \StringTok{"red"}\NormalTok{) }\SpecialCharTok{+} 
    \FunctionTok{ggtitle}\NormalTok{(}\StringTok{"Benchmarked to UN IGME"}\NormalTok{) }\SpecialCharTok{+} \FunctionTok{ylim}\NormalTok{(}\FunctionTok{c}\NormalTok{(}\DecValTok{0}\NormalTok{, }\FloatTok{0.22}\NormalTok{))}
\NormalTok{g1 }\SpecialCharTok{+}\NormalTok{ g2 }\SpecialCharTok{+}\NormalTok{ g3 }
\end{Highlighting}
\end{Shaded}

Finally, the subnational models can be fit in the same manner. Since the direct estimates are already on the yearly scale, we do not need to perform the transformation from periods to years. So we set \texttt{is.yearly} to FALSE to fit the period model directly.

\begin{Shaded}
\begin{Highlighting}[]
\NormalTok{fit.smooth.direct }\OtherTok{\textless{}{-}} \FunctionTok{smoothDirect}\NormalTok{(}\AttributeTok{data =}\NormalTok{ direct.comb.benchmark, }
                            \AttributeTok{Amat =}\NormalTok{ MalawiGraph, }
                            \AttributeTok{year.label =} \DecValTok{2000}\SpecialCharTok{:}\DecValTok{2019}\NormalTok{, }\AttributeTok{year.range =} \FunctionTok{c}\NormalTok{(}\DecValTok{2000}\NormalTok{, }\DecValTok{2019}\NormalTok{), }
                            \AttributeTok{time.model =} \StringTok{"rw2"}\NormalTok{, }\AttributeTok{m =} \DecValTok{1}\NormalTok{, }
                            \AttributeTok{type.st =} \DecValTok{4}\NormalTok{, }\AttributeTok{pc.alpha =} \FloatTok{0.05}\NormalTok{, }\AttributeTok{pc.u =} \DecValTok{1}\NormalTok{)}
\NormalTok{est.smooth.direct }\OtherTok{\textless{}{-}} \FunctionTok{getSmoothed}\NormalTok{(fit.smooth.direct)}
\end{Highlighting}
\end{Shaded}

\subsection{Cluster-level model estimation}

We now describe the fitting of the cluster level model using the two DHS surveys. With the created data frame, we first fit the national model with survey-year-specific HIV adjustment factors specified using \texttt{bias.adj} and \texttt{bias.adj.by} arguments. The adjustments are performed as offsets in the likelihood as described before. We also add a sum-to-zero survey effect term.

\begin{Shaded}
\begin{Highlighting}[]
\NormalTok{fit.bb.nat }\OtherTok{\textless{}{-}} \FunctionTok{smoothCluster}\NormalTok{(}\AttributeTok{data =}\NormalTok{ DHS.counts, }\AttributeTok{Amat =} \ConstantTok{NULL}\NormalTok{, }
                    \AttributeTok{family =} \StringTok{"betabinomial"}\NormalTok{, }\AttributeTok{year.label =} \DecValTok{2000}\SpecialCharTok{:}\DecValTok{2019}\NormalTok{, }
                    \AttributeTok{time.model =} \StringTok{"rw2"}\NormalTok{, }
                    \AttributeTok{bias.adj =}\NormalTok{ MalawiData}\SpecialCharTok{$}\NormalTok{HIV.yearly, }
                    \AttributeTok{bias.adj.by =} \FunctionTok{c}\NormalTok{(}\StringTok{"years"}\NormalTok{, }\StringTok{"survey"}\NormalTok{),}
                    \AttributeTok{survey.effect =} \ConstantTok{TRUE}\NormalTok{)}
\end{Highlighting}
\end{Shaded}

The \texttt{getSmoothed} function then produces estimates from the fitted cluster-level model. Similar to before, we take \texttt{nsim} draws of the posterior distribution to calculate the summaries of the U5MR estimates.

\begin{Shaded}
\begin{Highlighting}[]
\NormalTok{est.bb.nat }\OtherTok{\textless{}{-}} \FunctionTok{getSmoothed}\NormalTok{(fit.bb.nat, }\AttributeTok{nsim =} \DecValTok{1000}\NormalTok{, }\AttributeTok{save.draws =} \ConstantTok{TRUE}\NormalTok{) }
\end{Highlighting}
\end{Shaded}

In this example, no strata weights are provided and thus the overall estimates are empty. Given a data frame of strata proportions, we can rerun the \texttt{getSmoothed} function to re-aggregate the stratified estimates. The \texttt{save.draws} argument in the \texttt{getSmoothed} call allows the raw posterior draws to be returned as part of the output object. This can be helpful in such situations, as posterior draws already computed can be inserted into new \texttt{getSmoothed} calls using the \texttt{draws} argument to avoid resampling again.

\begin{figure}[!ht]
\includegraphics[width=.8\textwidth,]{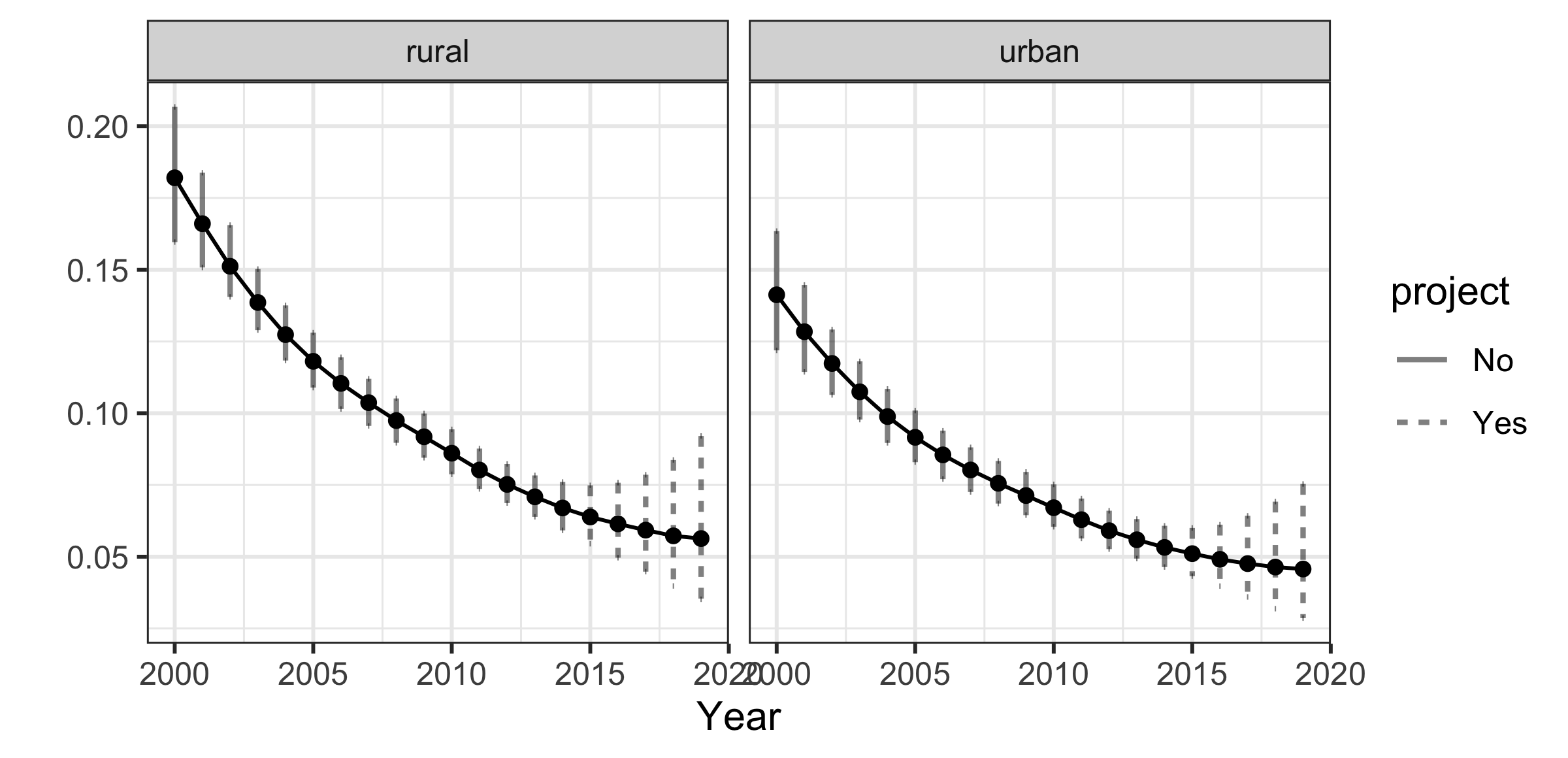} \caption{National estimated rural and urban  U5MR in Malawi using DHS from 2010 and 2015--2016.}\label{fig:bb-national-4}
\end{figure}

Figure \ref{fig:bb-national-4} shows the national estimates of U5MR in Malawi for urban and rural stratum respectively using the cluster-level model.

\begin{Shaded}
\begin{Highlighting}[]
\FunctionTok{plot}\NormalTok{(est.bb.nat}\SpecialCharTok{$}\NormalTok{stratified, }\AttributeTok{is.subnational=}\ConstantTok{FALSE}\NormalTok{, }\AttributeTok{proj.year =} \DecValTok{2016}\NormalTok{) }\SpecialCharTok{+} \FunctionTok{facet\_wrap}\NormalTok{(}\SpecialCharTok{\textasciitilde{}}\NormalTok{strata) }
\end{Highlighting}
\end{Shaded}

We can also fit the subnational model in the same fashion. Additional analysis can be similarly carried out as in the previous example with simulated data. For the cluster-level model, benchmarking to national estimates is also implemented in the package using the procedure described in {[}@okonek2022computationally{]} with additional information on population fractions by region. We refer readers to the package vignette for more details.

\begin{Shaded}
\begin{Highlighting}[]
\NormalTok{fit.bb }\OtherTok{\textless{}{-}} \FunctionTok{smoothCluster}\NormalTok{(}\AttributeTok{data =}\NormalTok{ DHS.counts, }\AttributeTok{Amat =}\NormalTok{ MalawiGraph, }
                    \AttributeTok{family =} \StringTok{"betabinomial"}\NormalTok{, }
                    \AttributeTok{year.label =} \DecValTok{2000}\SpecialCharTok{:}\DecValTok{2019}\NormalTok{, }
                    \AttributeTok{time.model =} \StringTok{"rw2"}\NormalTok{, }\AttributeTok{st.time.model =} \StringTok{"ar1"}\NormalTok{,}
                    \AttributeTok{pc.st.slope.u =} \DecValTok{2}\NormalTok{, }
                    \AttributeTok{pc.st.slope.alpha =} \FloatTok{0.1}\NormalTok{,}
                    \AttributeTok{bias.adj =}\NormalTok{ MalawiData}\SpecialCharTok{$}\NormalTok{HIV.yearly, }
                    \AttributeTok{bias.adj.by =} \FunctionTok{c}\NormalTok{(}\StringTok{"years"}\NormalTok{, }\StringTok{"survey"}\NormalTok{),}
                    \AttributeTok{survey.effect =} \ConstantTok{TRUE}\NormalTok{, }
                    \AttributeTok{strata.time.effect =} \ConstantTok{TRUE}\NormalTok{)}
\NormalTok{est.bb }\OtherTok{\textless{}{-}}  \FunctionTok{getSmoothed}\NormalTok{(fit.bb, }\AttributeTok{nsim =} \DecValTok{1000}\NormalTok{, }\AttributeTok{save.draws =} \ConstantTok{TRUE}\NormalTok{) }
\end{Highlighting}
\end{Shaded}

The U5MR by strata can be visualized directly. Notice that in order to produce the overall estimates by region and time, additional information on population fractions in urban/rural is necessary, similar to the analysis conducted in the main paper using simulated data. For more details on obtaining such factions, we refer readers to {[}@fuglstad2021two{]} and {[}@wu2021spatial{]}.

\begin{Shaded}
\begin{Highlighting}[]
\FunctionTok{plot}\NormalTok{(est.bb}\SpecialCharTok{$}\NormalTok{stratified) }\SpecialCharTok{+} \FunctionTok{facet\_wrap}\NormalTok{(}\SpecialCharTok{\textasciitilde{}}\NormalTok{strata)}
\end{Highlighting}
\end{Shaded}

\begin{figure}[!ht]
\includegraphics[width=\textwidth,]{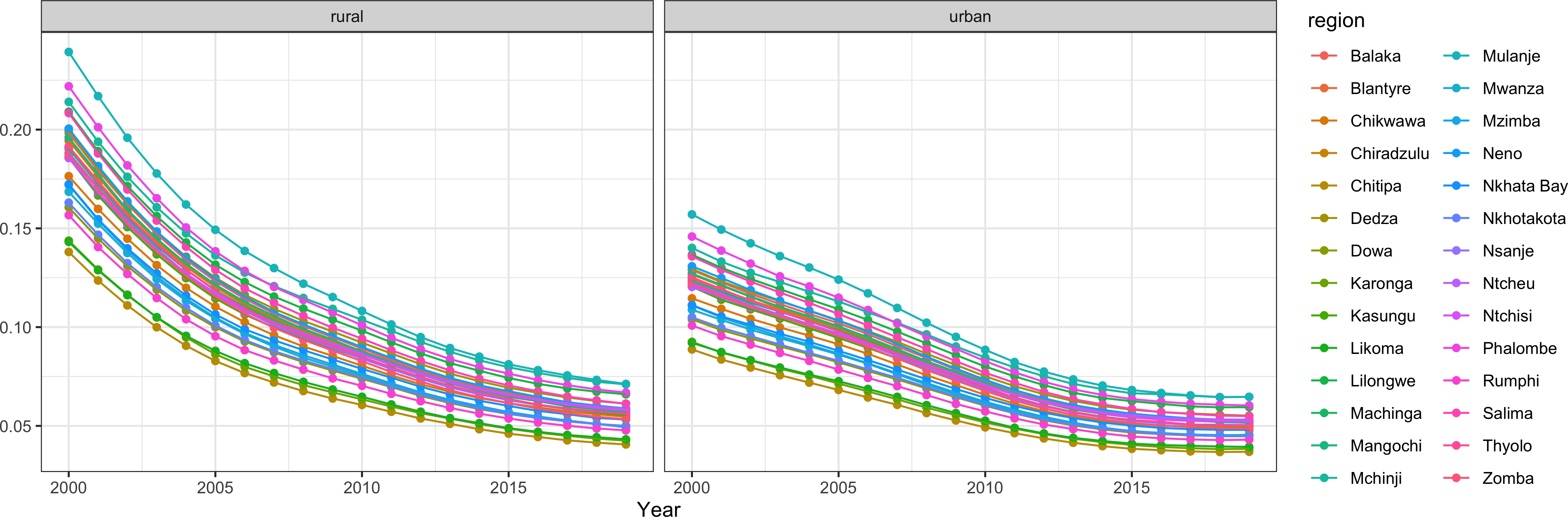} \caption{Subnational estimated rural and urban  U5MR in Malawi using DHS from 2010 and 2015--2016.}\label{fig:bb-sub-2}
\end{figure}

\bibliographystyle{apalike}
\bibliography{SUMMER-RJ.bib}

\end{document}